\documentstyle{article}
\newfont{\footsc}{cmcsc10 at 8truept}
\newfont{\footbf}{cmbx10 at 8truept}
\newfont{\footrm}{cmr10 at 10truept}

\begin{document}

\centerline{ \Large \bf Characteristic polynomials of random
Hermitian matrices} \centerline{\Large \bf  and
Duistermaat-Heckman localisation} \centerline{\Large \bf on
non-compact K\"{a}hler manifolds}

\vskip 0.5cm \centerline{ \large \bf Yan V Fyodorov and Eugene
Strahov}

\centerline{Department of Mathematical Sciences, Brunel
University} \centerline{Uxbridge, UB8 3PH, United Kingdom}
\centerline{\small \texttt{Yan.Fyodorov@brunel.ac.uk}}
\centerline{\small \texttt{Eugene.Strahov@brunel.ac.uk}} \vskip
0.3cm

\begin{abstract}
We reconsider the problem of calculating a
 general spectral correlation function containing an arbitrary number of
 products and ratios of characteristic polynomials for a $N\times N$
random matrix taken from the Gaussian Unitary Ensemble (GUE).
 Deviating from the standard "supersymmetry" approach, we
integrate out Grassmann variables at the early stage and
 circumvent the use of the Hubbard-Stratonovich transformation
in the "bosonic" sector. The method, suggested recently by one of us \cite{I},
 is shown to be capable of calculation when reinforced with a
generalization of the Itzykson-Zuber integral to a non-compact integration manifold.
We arrive to such a generalisation by discussing the
Duistermaat-Heckman localization principle for integrals
over non-compact homogeneous K\"{a}hler manifolds.
 In the limit of large $N$ the asymptotic expression
for the correlation function reproduces the result outlined earlier by
Andreev and Simons \cite{AS}.

\end{abstract}

\section{Introduction}

Recently there was an outburst of research activity related to
investigating the moments and correlation functions of
characteristic polynomials $Z_N(\mu)=\det{\left(\mu {\bf
1}_N-\hat{H}\right)}$ for random $N\times N$ matrices $H$ of
various types. Those studies were motivated by hope to
relate statistics of zeroes of the Riemann zeta function to
that of eigenvalues of large random matrices
\cite{zeta1,zeta2,zeta3,BH,web,BK},
 as well as by numerous applications of spectral determinants
in the theory of quantum chaotic and disordered systems
\cite{Gan,GK,U1,U2,F,MN,KM,LY,AS,FK99,AGG,TS,Ket,I,ZN}, quantum
chromodynamics \cite{DN},
 and relations to interesting combinatorial problems \cite{E}.

There are several analytical techniques for
 dealing with the integer moments (positive or negative)
of characteristic polynomials. Their applicability varies with the
nature of the underlying random matrix ensemble. For unitary
random matrices ( the so-called "unitary circular ensemble") one
can either relate evaluation of the moments to the
 Selberg-type integral \cite{zeta1,zeta2}, or rely
upon variants of the character expansion, directly \cite{E} or
indirectly \cite{Zirn}. Another well-studied case relates to the
ensembles of Hermitian random matrices characterized by unitary
invariant probability measures. That invariance allows for {\it
positive} integer moments of the characteristic polynomials
 to be evaluated by methods resorting to  orthogonal polynomials
 \cite{BH,MN}.
The particular case of the
 Gaussian measure  can be studied very efficiently
 by the following procedure. First,
one represents each of the characteristic polynomials as a Gaussian
integral over anticommuting (Grassmann) variables. This allows
  to average the resulting expressions immediately.
At the next step one employs the so-called Hubbard-Stratonovich
transformation combined with the subsequent exploitation of the
Itzykson-Zuber -Harich-Chandra integral \cite{IZ,HC}:
\begin{equation}\label{IZ}
\int\limits_{g\in
\sf{U(N)}/\sf{T}}d\mu(g)\exp\left[i\;\mbox{Tr}\left(XgYg^{\dag}\right)
\right]= \mbox{const}\;\frac{\mbox{det}\left[\exp(ix^l
y^k)\right]}{\triangle(X)\triangle(Y)}\left.\right|_{1\leq l,k\leq
N}
\end{equation}
Here $X$, $Y$ are diagonal matrices with eigenvalues $x^l,y^k$
correspondingly , where indices $k$ and $l$ take values from $1$
to $N$. $\triangle(X),\triangle(Y)$ stand for the Vandermonde
determinants and $\sf{T}={\sf{U}}(1)\times\cdots\times
{\sf{U}}(1)$ is the maximal torus of the group $\sf{U(N)}$ . A
detailed outline of the method can be found, e.g. in \cite{KM}.

The evaluation of the {\it negative} integer moments of
characteristic polynomials turns out to be more tricky as care
must be taken to avoid divergences and account for presence of
poles \cite{I,BK}. The standard way goes back to the work by
Sch\"{a}fer and Wegner \cite{SW} and relies upon representing the
(regularized) inverse determinants as the Gaussian integrals over
commuting complex variables. This method then  exploits an
extension of the Hubbard-Stratonovich transformation to a
non-trivial manifold with inherent "hyperbolic" structure. An
alternative variant of the method was suggested in a recent paper
by one of the present authors \cite{I}, referred as [I]
henceforth. The latter work contains a detailed discussion of the
problem as well as many related references.

A much more general correlation function of characteristic
polynomials is one combining presence of both positive and
negative integer moments.
Such a correlation function contains a
very detailed information about spectra of random matrices and
thus it is  most important for applications in physics.
 Correlation functions of that type are also interesting for
the sake of comparison with more refined conjectures on the
behavior of the Riemann zeta-function\cite{Farmer}.

 The standard
technique in that case naturally combines Gaussian integrals over
commuting and anticommuting variables and the subsequent
Hubbard-Stratonovich transformation of usual and "hyperbolic"
nature. The method is known in the literature as the supermatrix
(or "supersymmetry") approach pioneered by Efetov \cite{Efetov}
in the theory of disordered systems, and taken over to random
matrices by Verbaarschot and Zirnbauer \cite{VZ}.

It should work, in principle, for the general case but technically
any general calculation beyond the two-point correlation function proved
to be extremely difficult. The main problem is related to the
so-called "anomalous", or "boundary" terms arising
when changing variables in the superintegrals. Those anomalies
can be traced back to the admixture of nilpotent terms ( those
containing even number of grassmannian factors)
to the usual commuting variables. All such anomalous
terms were classified by Rothstein\cite{anom} in a general form.
 However, to write down their contribution explicitly
in a specific parameterization proves to be a very daunting job.
The latter fact makes the standard supersymmetric calculations
usually impractical beyond a few lower-order correlation
functions. A notable exception is the general many-point
correlation function of spectral densities (see works by
Zirnbauer\cite{Zirna} and Szabo\cite{Szabo1}). In the latter case
the boundary terms do not contribute
  and the final result is provided solely by the "bulk"
integral which is evaluated by standard methods. At the same
time our main object
of interest - the correlator of spectral determinants - contains
anomalous terms on equal footing with the bulk contribution.

A modification of the supermatrix
method was suggested by Guhr \cite{Guhr}. It relied upon a
generalization of the Itzykson-Zuber integral to a unitary
supergroup and allowed one to go beyond the two-point function for
the case of Gaussian Unitary Ensemble (GUE).
A few years ago Andreev and Simons  presented an asymptotic formula for a
correlation function containing both products and ratios of
characteristic polynomials for GUE matrices. In their short
communication\cite{AS} they claimed that Guhr's method equipped
with the further extension of the Itzykson-Zuber integral to a
{\it pseudo}unitary supergroup solved the problem. The authors
indicated that the arguments behind that generalization were
similar in spirit to those by Guhr, but "technically involved".
They promised to present details of the method "in a longer paper"
which, unfortunately, never appeared. Let us mention that
the role of boundary contributions in the Guhr's method seems to be
not clearly discussed in the literature.

In the present paper we show that the method suggested in [I] enables one
to calculate the general correlation function of integer moments
of GUE characteristic polynomials. In fact, we integrate out
Grassmann variables at a very early stage thus seriously departing from the
general spirit of supersymmetry. One of the advantages is that no
anomalous terms can ever arise along such a route.

The calculation required, however, the knowledge of an
analogue of the Itzykson-Zuber type integral over {\it
non-compact} manifolds related to pseudo-unitary groups. A
pseudo-unitary group $\sf{G}$ is defined by the following
conditions imposed on the group elements:
 \begin{equation}\label{pseudounitarygroup}
g^{\dag}\lambda g=\lambda,\;\;\;\forall g\in\sf{G}
\end{equation}
Here  $\lambda$ is a diagonal matrix with the  elements $\pm 1$ in
a matrix representation.

In the mathematical literature the integrals of the Itzykson-Zuber
type over semi-simple Lie groups are interpreted as Fourier
transforms over adjoint orbits $gXg^{-1}, g\in {\sf{G}}$ and
formulas for such integrals are known (see the works of Rossmann
\cite{rossmann}, Berline and Vergne \cite{berline}, Prato and
Wu\cite {prato}, Paradan \cite{paradan}) . However the derivation
of the Itzykson-Zuber type integrals over non-compact groups in a
manner accessible to physicists was not presented before to the
best of our knowledge.

A particular case of non-compact extension of the Itzykson-Zuber
integral was presented recently in the Appendix C of [I].
 The method of calculation, however, relied upon a specific
 parameterization of the integration manifold
and seemed hardly applicable to a general non-compact case.

The standard procedure of the derivation of the Itzykson-Zuber type integrals
 is to employ the diffusion equation arguments.
This method goes back to the original paper by Itzykson and Zuber
\cite{IZ} and was also used by Guhr \cite{Guhr}, and, apparently by
Andreev and Simons\cite{AS}. We provide such
derivation for the Itzykson-Zuber type integral over the pseudo-unitary
group in our Appendix A.

In this paper we consider in detail a different approach based on
the Duistermaat-Heckman localisation principle \cite{DH} outlined
in the context of random matrices  by Zirnbauer \cite{Zirn}.
Indeed, it is well known that the Itzykson-Zuber integral is a
representative of the family of integrals of the form $\int
\Omega^m\exp{iH}$
 going over a $2m-$ dimensional phase space.  These integrals are
  semiclassically exact
provided the Hamiltonian $H$ is "localisable". In the case of
compact phase spaces localisability is equivalent to the condition
for the phase flow generated by the Hamiltonian $H$ to preserve
the Riemannian metric of the phase space.  In particular,
considering the manifold $\sf{U(N)}/\sf{T}$ as a phase space,  the
Itzykson-Zuber Hamiltonian
\begin{equation}\label{izham}
H(X,Y,g)=\mbox{Tr}\left(XgYg^{\dag}\right)
\end{equation}
appears to be localisable in the sense of  Duistermaat and
Heckman.

From the geometrical point of view  the underlying phase space
$\sf{U(N)}/\sf{T}$ of the Itzykson-Zuber formula belongs to the
family of compact flag manifolds . The flag manifolds are
K\"{a}hler homogeneous spaces of the type $\sf{G}/\sf{H}$, where
$\sf{G}$ is called a transformation group and $\sf{H}$ is the
centralizer of a sub-torus of ${\sf{T}}$ in the group G, i.e.
${\sf{H}}=\left\{g\in {\sf{G}}|g^{-1}{\sf{T}}_0
g={\sf{T}}_0\right\}$, where ${\sf{T}}_0$ is a sub-torus of
${\sf{T}}$. The compact flag manifolds were considered in details
by Picken \cite{picken} in the context of the Duistermaat-Heckman
formula. Picken has obtained an explicit expression for the set of
localizable Hamiltonians on a compact flag manifold. Then the fact
that the Itzykson-Zuber Hamiltonian is  localisable follows as a
particular case $(\sf{G}=U(N), \sf{H}=U(1)\times\ldots\times
U(1))$.

The possibility to extend the Duistermaat-Heckman localisation
principle  to non-compact symplectic spaces was first discussed  by
Prato and Wu \cite{prato}. The case of the non-compact counterpart
of the space $CP^{N}$ was considered  by  Fujii and Funahashi
\cite{fujii} . However the calculations done in [I] suggest
  to  concentrate on more general homogeneous
 non-compact symplectic manifolds. We show that an
 extension of the Duistermaat-Heckman formula is possible when the homogeneous
manifold under considerations, $\sf{G}/\sf{H}$ is a non-compact
\textit{K\"ahler} manifold. The key feature of such manifolds
enabling one to apply the Duistermaat-Heckman theorem is that
their invariant metrics can be chosen to be sign-definite. In
turn, the review paper by Bordemann, Forger and Romer
\cite{bordemann} provides conditions under which a homogeneous
manifold turns out to be K\"ahlerian.  The conditions can be
summarized as follows. Assume that the group $\sf{G}$ is a
connected and semisimple and  the subgroup $\sf{H}$ is a compact
centralizer of a torus in $\sf{G}$. Then
\begin{enumerate}
  \item If $\sf{M}=\sf{G/H}$ is  compact, $\sf{M}$ is a K\"ahler
  manifold.
  \item Let $\sf{M=G/H}$ be non-compact and $\sf{L}$ be  the maximal
  compact subgroup of $\sf{G}$ containing $\sf{H}$. Then
  $\sf{M}=\sf{G/H}$ is a K\"ahler manifold if and only if
  $\sf{G/L}$ is a Hermitian symmetric space.
\end{enumerate}
In particular, the manifold $\sf{U(n_1,n_2)}/\sf{T}$ appears to be
K\"ahlerian in this case. Indeed, the maximal compact subgroup of
$\sf{U(n_1,n_2)}$ containing $\sf{H}=\sf{T}$ is
$\sf{L}=\sf{U(n_1)\times U(n_2)}$. Since
$\sf{U(n_1,n_2)/U(n_1)\times U(n_2)}$ is a Hermitian symmetric
space of the type $\textsf{AIII}$ (see Helgason \cite{helgason},
p.354) the manifold $\sf{U(n_1,n_2)}/\sf{T}$ is a K\"ahler
manifold as follows from the criterions presented above.

Our general construction is then exploited to derive  the
Itzykson-Zuber type integral over a pseudo-unitary group
\begin{eqnarray}\label{IZpseudo}
\int\limits_{g\in \sf{U(n_1, n_2
)}/\sf{T}}d\mu(g)\exp\left[i\;\mbox{Tr}\left(XgYg^{-1}\right)\right]=
\nonumber\qquad\qquad\\
\mbox{const}\;\frac{\mbox{det}\left[\exp(ix^l
y^k)\right]\left.\right|_{1\leq l,k\leq
n_1}\mbox{det}\left[\exp(ix^l y^k)\right]\left.\right|_{n_1+1\leq
l,k\leq n_1+n_2 }}{\triangle(X)\triangle(Y)}
\end{eqnarray}
It then can be seen  that the remarkable formula of Harish-Chandra
\cite{HC} remains valid also for the non-compact group
$\sf{U(n_1,n_2)}$. Indeed, the righthand part of
Eq.(\ref{IZpseudo}) can be rewritten as
\begin{eqnarray}\label{IZpseudo0}
\frac{\mbox{const}}{\prod\limits_{\alpha
>0}\alpha(X)\prod\limits_{\alpha>0}\alpha(Y)}
\sum\limits_{w\in\;
{\sf{W}}}(-)^{|w|}\exp\left[i\;\mbox{Tr}\left(Xw(Y)\right)\right]
\nonumber
\end{eqnarray}
where $\sf{W}$ is the Weyl group corresponding to
$\sf{U(n_1,n_2)}$, i.e.  $\sf{W}=S_{n_1}\times S_{n_2}$. The
$\alpha(X)$ is a root corresponding to a Cartan subalgebra element
$X$, $w(Y)= wYw^{-1}$ and $|w|$ denotes the parity of $w$.

  The structure of the
paper is as follows. In sections 2 and 3 we provide a necessary
background information on K\"{a}hler geometry and discuss the
Duistermaat-Heckman localisation on non-compact K\"ahler
manifolds. Then in section 4 we give an account of (a refined
version of) the method suggested in [I] for calculating the
general correlation function of characteristic polynomials for the
GUE matrices. In section 5 we analyse the derived
 matrix integral representation. For this purpose we use  the
formulae Eqs.(\ref{IZ}) and (\ref{IZpseudo})
 and evaluate the remaining integrals by
the saddle-point method in the limit $N\to \infty$. The open questions
are summarized in the Conclusions. Technical details are presented
in the appendices.

\section{Basic properties of K\"{a}hler manifolds}
In this section we show how the Duistermaat-Heckman \cite{DH}
localisation principle
 can be reformulated
for the K\"ahlerian dynamical systems. These are dynamical systems
whose  phase spaces are (simply connected) homogeneous K\"{a}hler
manifolds ( for definitions and basic properties of homogeneous
K\"{a}hler manifolds see, for example, Kobayashi and Nomizu
\cite{kobayashi}).

Throughout the paper we use a complex parameterization on flag
manifolds which is introduced following Borel's method
\cite{borel}. A detailed exposition of the method can be found in
the papers by Bar-Moshe and Marinov \cite{marinov1},
\cite{marinov2}. Below we provide a concise description of the main
features of the underlying structures.

Given a semi-simple Lie group  $\sf{G}$ we introduce the canonical
Cartan-Weil basis for the corresponding complex semi-simple Lie
algebra \textbf{g}: $\left\{\tau_a\right\}=\left\{h_j,e_{\pm
q}\right\}$. Here $a=1,2,\ldots ,n=\mbox{dim}\;\bf{g}$;
$j=1,2,\ldots ,r=\mbox{rank}\; \textbf{g}$, and
$\left\{q\right\}\in \triangle^{+}_{\textbf{g}}$ are the positive
roots of the Lie algebra \textbf{g}.  The complex parameters which
are introduced in the flag manifold $\sf{G}/\sf{H}$ correspond to
the positive roots of the Lie algebra $\bf{g}$. When the subgroup
$\sf{H}$ of the group $\sf{G}$ is a maximal torus,
$\sf{H}=\sf{T}$, any element $g(z,\bar z)$ of the coset space
$\sf{G/T}$ has the following decomposition:
\begin{eqnarray}\label{gcdecomposition}
 g(z,\bar z)=u(z)p(z,\bar z)\qquad\qquad\qquad\qquad\qquad\qquad\nonumber \\
u(z)=\exp\left(\sum\limits_{q\in\triangle^{+}_{\textbf{g}}}z^{q}e_q\right),\;\;\;\;
p(z,\bar z)
=\exp\left(\sum\limits_{q\in\triangle^{+}_{\textbf{g}}}y^{q}
(z,\bar z)e_{-q}\right)\exp\left(\sum\limits_{j=1}^{r}k^j(z,\bar
z)h_j\right)
\end{eqnarray}
where we used the bar to denote the complex conjugation.

To find an explicit expression for $y^{q}$ and $k^j$
 as functions of the complex
coordinates $z^{q}, z^{\bar q}\equiv\overline{z^q}$, one has to
exploit the condition $g^{\dag}(z,\bar z)=g^{-1}(z,\bar z)$ when
the group
 $\sf{G}$ is unitary, or the constraint
(\ref{pseudounitarygroup}) for the case of a pseudo-unitary group.
The element $g(z,\bar z)=u(z)p(z,\bar z)$ constructed in this way
represents a point with the coordinates $(z^q, z^{\bar q})$ on the
flag manifold $\sf{G}/\sf{H}$. To provide a reader with a simple
but informative example  we show in the Appendix B an application
of the general principles outlined in this section for the case of
the compact manifold $\sf{U(2)/U(1)\times U(1)}$ and its
non-compact counterpart $\sf{U(1,1)/U(1)\times U(1)}$.

Let us note that a complex parameterization introduced above can
be looked at as a convenient way of describing the action of the
transformation group ${\sf{G}}$ on its flag manifold $\sf{G/H}$.
Indeed, for any $g\in \sf{G}$ the (unique) decomposition
$gu(z)=u(gz)p(z,g)$ allows one to find $gz$ and thus to determine
the (holomorphic) action of the element $g$ of the transformation
group $\sf{G}$ on a point of the flag manifold $\sf{G/H}$.

The homogeneous K\"{a}hler manifold $\sf{M}$ comes with the
K\"{a}hler potential $K(z,\bar z)$, which is a scalar function
defined on any open neighborhood of $\sf{M}$ with local complex
coordinates $z^{\alpha}$, $(\alpha=1,2,\ldots
,m=\mbox{dim}_C\sf{M})$. When  a group $\sf{G}$ acts
holomorphically on $\sf{M}$,
\begin{equation}
z\rightarrow gz, \forall g\in\sf{G}
\end{equation}
the K\"{a}hler potential $K(z,\bar z)$ is transformed as
\begin{equation}\label{ktransform}
K(z,\bar z)\rightarrow K(gz,\overline{gz})=K(z,\bar
z)+\Phi_g(z)+\overline{\Phi_g(z)},\;\; g\in\sf{G}
\end{equation}
where $\Phi_g(z)$ is a holomorphic function of $z$. Once the
K\"ahler potential is provided, the exact (1,1) differential
 form on the
K\"ahler manifold is introduced as follows: \footnote{Throughout
the paper we assume tensor notations,
 i.e. summation over repeating indices}
\begin{equation}\label{oneoneform}
\Omega=\omega_{\alpha\bar\beta}\left(z,\bar
z\right)dz^{\alpha}\wedge dz^{\bar\beta},\;\;
\omega_{\alpha\bar\beta}\left(z,\bar z\right)=-\frac{1}{2\pi
i}\;\partial_{\alpha}\partial_{\bar\beta}K(z,\bar z)
\end{equation}
where the factor $-1/(2\pi i)$ is chosen for  convenience.
Equations Eqs.(\ref{oneoneform}), (\ref{ktransform}) show that the
(1,1) form $\Omega$ is invariant under the holomorphic action of
the group $\sf{G}$ on the homogeneous K\"{a}hler manifold
$\sf{M}$. When the phase space (K\"{a}hler manifold $\sf{M}$) and
the (1,1) form are specified, the classical mechanics is defined
by the Poisson brackets for any two smooth functions $F_1(z,\bar
z)$ and $F_2 (z,\bar z)$:
\begin{equation}\label{poisson}
\left\{F_1(z,\bar z),F_2(z,\bar z)\right\}_{P.B.}=
w^{\alpha\bar\beta}(z,\bar z)\left(\partial_{\alpha}F_1(z,\bar
z)\partial_{\bar\beta}F_2(z,\bar z)-\partial_{\alpha}F_2(z,\bar
z)\partial_{\bar\beta}F_1(z,\bar z)\right)
\end{equation}
where the antisymmetric field $w^{\alpha\bar\beta}(z,\bar z)$ is
inverse to $w_{\alpha\bar\beta}(z,\bar z)$, i.e.
\begin{equation}\label{inverse}
w^{\alpha\bar\beta}(z,\bar z)w_{\nu\bar\beta}(z,\bar
z)=\delta^{\alpha}_{\nu},\;\;\;w^{\nu\bar\beta}(z,\bar
z)w_{\nu\bar\alpha}(z,\bar z)=\delta^{\bar\beta}_{\bar\alpha}
\end{equation}
Provided that a Hamiltonian $H(z,\bar z)$ is given , the equations
of motion on the K\"{a}hler homogeneous manifold $\sf{M}$ can be
written in terms of the Poisson brackets,
\begin{equation}\label{pbracket} dF(z,\bar
z)/dt=\left\{F,H\right\}_{P.B.}
\end{equation}
From Eqs.(\ref{ktransform})-(\ref{pbracket}) we then obtain
a time evolution of the complex coordinates:
\begin{equation}\label{zevolution}
\dot{z}^{\alpha}=w^{\alpha\bar\beta}(z,\bar
z)\partial_{\bar\beta}H, \;\;
\dot{z}^{\bar\beta}=-w^{\alpha\bar\beta}(z,\bar
z)\partial_{\alpha}H
\end{equation}

In the formulation of the Duistermaat-Heckman localisation
principle
 one uses an important notion of the Hamiltonian phase flow
 defined on the homogeneous K\"ahler manifold $\sf{M}$
 as a one-parameter group
$g(t)$ of diffeomorphisms $\sf{M}\rightarrow\sf{M}$ (see, for
example, Arnold \cite{arnold}):
\begin{eqnarray}\label{phaseflow}
d/dt|_{t=0}\left(g(t)\cdot
z^{\alpha}\right)=w^{\alpha\bar\beta}(z,\bar z)
\partial_{\bar\beta}H, \;\; d/dt|_{t=0}\left(g(t)\cdot
z^{\bar\beta}\right)=-w^{\alpha\bar\beta}(z,\bar z)
\partial_{\alpha}H
\end{eqnarray}
From the above definition and Eqs.(\ref{ktransform}),
(\ref{oneoneform}) it follows that the K\"{a}hler metric remains
invariant under the holomorphic phase flow.

 Let us now represent an
arbitrary element of the Lie group $\sf{G}$ acting holomorphically
on the K\"{a}hler manifold $\sf{M}$ by the set of cartesian
coordinates:
\begin{equation}\label{groupcartesian}
g(\xi)=\exp\left(\xi^a\tau_a\right),\;\;\; a=1,2,\ldots
,n=\mbox{dim}\;{\bf{g}}
\end{equation}
where $\tau_a$ are basis elements of the Lie algebra ${\bf g}$ of
the group $\sf{G}$ satisfying the commutation relations
\begin{equation}\label{algebracommutations}
\left[\tau_a,\tau_b\right]=f_{ab}^{c}\tau_c,
\end{equation}
with $f_{ab}^c$ being the structure constants of the Lie algebra
$\bf{g}$.

The Lie group $G$ is itself a homogeneous space so the left action
of the group on itself induces vector fields:
\begin{equation}\label{vfieldsgroup}
\tau_a\rightarrow
\textsf{D}_a(\xi)=\emph{L}_a^b(\xi)\partial^{\xi}_b
\end{equation}
where $\partial^{\xi}_b$ denotes a partial derivative  with
respect to the parameter $\xi^b$.

 The holomorphic action of the Lie group ${\sf{G}}$ on the K\"ahler manifold $\sf{M}$
$z\in{\sf{M}}\rightarrow gz\in {\sf{M}}, \forall g\in {\sf{G}}$
 induces the vector fields
  of the form
\begin{equation}\label{vectorfieldsmanifold}
\nabla_a(z,\bar z)=\kappa_a^{\alpha}(z)
\partial_{\alpha}+\kappa_a^{\bar\alpha}(\bar
z)\partial_{\bar\alpha}
\end{equation}
where the fields $\kappa_a^{\alpha}(z)$ are expressed by the
induced vector fields on the group:
\begin{equation}\label{killings}
\kappa_a^{\alpha}(z)=\textsf{D}_a(\xi)\left(g(\xi)z\right)^{\alpha}|_{\xi=0}
\end{equation}
The conjugate fields $\kappa_a^{\bar\alpha}(\bar z)$ are defined
similarly. Once the $\nabla_a(z,\bar z)$ are induced vector fields
they also satisfy  the commutation relations
(\ref{algebracommutations}) of the Lie algebra $\bf{g}$.

Introduce now the linear operators $i_a$ acting on the
differential form $\Omega$ as:
\begin{equation}
i_a\Omega=w_{\alpha\bar\beta}(z,\bar
z)\kappa^{\alpha}_a(z)dz^{\bar\beta}-w_{\alpha\bar\beta}(z,\bar
z)\kappa^{\bar\beta}_a(\bar z)dz^{\alpha}
\end{equation}
The linear operators $i_a$ are related to the Lie derivatives
${\bf{\textsc{L}}}_{\nabla_a}$ corresponding to the vector fields
$\nabla_a(z,\bar z)$:
\begin{equation}
{\bf{\textsc{L}}}_{\nabla_a}=di_a+i_ad
\end{equation}
where $d$ stands for the usual external derivative operator acting
on forms.
 The invariance
of the (1,1) K\"{a}hler form $\Omega $, Eq. (\ref{oneoneform})
under the holomorphic action of the Lie group $\sf{G}$ can be
represented by the condition
\begin{equation}\label{LieDerivative11Form}
{\bf{\textsc{L}}}_{\nabla_a}\Omega=0,\;\;\; \forall a=1,2,\ldots
,n=\mbox{dim}\;\bf{g}
\end{equation}
Since the K\"{a}hler form $\Omega $ is closed: $d\Omega=0$ the
condition Eq.(\ref{LieDerivative11Form}) can be rewritten as
\begin{equation}\label{diaOmega}
di_a\Omega=0
\end{equation}
As the K\"{a}hler manifold $\sf{M}$ is assumed to be simply
connected, and in simply connected spaces all closed forms are
exact: $d\Omega=0 \Rightarrow \Omega= d\omega$, the equation above
implies the existence of $n=\mbox{dim}\;\bf{g}$ functions
$T_a(z,\bar z)$ (unique up to an additive constant) such that
\begin{equation}\label{iaOmegaequalT}
i_a\Omega=(2\pi i)^{-1}dT_a
\end{equation}

In terms of the local complex coordinates on the K\"{a}hler
manifold $\sf{M}$ differential equations (\ref{iaOmegaequalT})
acquire the form:
\begin{equation}\label{difmoment}
\partial_{\alpha}T_a(z,\bar z)=-2\pi i\kappa^{\bar\beta}_a(\bar
z)w_{\alpha\bar\beta}(z,\bar z),\;\;
\partial_{\bar\beta}T_a(z,\bar z)=2\pi i\kappa^{\alpha}_a(z)w_{\alpha\bar\beta}(z,\bar z)
\end{equation}
An immediate consequence of these equations is
\begin{equation}\label{difmoment1}
\omega^{\alpha \bar\beta}(z,\bar z)\partial_{\alpha}T_a(z,\bar z)
=-2\pi i\kappa^{\bar\beta}_a(\bar z),\;\;\;\omega^{\alpha
\bar\beta}(z,\bar z)\partial_{\bar\beta}T_a(z,\bar z) =2\pi
i\kappa^{\alpha}_a(z)
\end{equation}
which, in turn, implies the relations:
\begin{equation}\label{difmoment2}
\partial_{\bar\beta}\left(\omega^{\mu \bar\nu}(z,\bar z)
\partial_{\bar \nu}T_a(z,\bar z)\right)
=0,\;\;\; \partial_{\alpha}\left(\omega^{\mu \bar\nu}(z,\bar z)
\partial_{\mu}T_a(z,\bar z)\right)=0
\end{equation}

The functions $T_a(z,\bar z)$ are called equivariant momentum
maps, and we will see that they play an important role in
providing the relation between the K\"{a}hler geometry and the
Duistermaat-Heckman localisation principle. The explicit
construction of the equivariant momentum maps $T_a(z,\bar z)$ for
the compact flag manifolds was performed by Bar-Moshe and
Marinov\cite{marinov1}\cite{marinov2} and will be discussed later
on in the present paper.

\subsection{Duistermaat-Heckman localisation and
equivariant momentum maps}

Now we will obtain the conditions under which a Hamiltonian
$H(z,\bar z)$ on the K\"ahler manifold can be localisable in the
sense of the Duistermaat-Heckman theorem. In this section we adopt
the method developed previously by Bismut\cite{bismut},
Witten\cite{witten}, Zirnbauer\cite{Zirn} and apply it for the
particular case of the K\"{a}hler manifolds.

We consider the integral
\begin{equation}\label{integral}
I=\int\limits_{\sf{M}}\Omega^m\exp\left(iH(z,\bar z)\right)
\end{equation}
where $\sf{M}$ is a K\"{a}hler manifold on which the complex
parameterization is introduced. When $\sf{M}$ is a flag manifold
it has a complex parameterization as discussed above, and
 the (1,1) form $\Omega$ is given by Eq.(\ref{oneoneform}), with $m$
being the complex dimension of the
manifold $\sf{M}$, $\mbox{dim}_C{\sf{M}}=m$. Let us introduce $2m$
anticommuting variables $\left(\xi^{\alpha},
\xi^{\bar\alpha}\right)$ that are counterparts to commuting
complex coordinates $(z^{\alpha}, z^{\bar\alpha})$ of the manifold
$\sf{M}$.  Integral (\ref{integral}) can be rewritten  as that
with a flat integration measure:
\begin{equation}\label{intflat}
I=\int\prod\limits_{\alpha,\bar\alpha=1}^{m}
dz^{\alpha}dz^{\bar\alpha}\prod\limits_{\beta,\bar\beta=1}^{m}
d\xi^{\beta}d\xi^{\bar\beta}\exp{S(z,\bar z,\xi,\bar\xi)}
\end{equation}
where
\begin{equation}\label{action}
S(z,\bar z,\xi,\bar\xi)=iH(z,\bar z)-w_{\alpha\bar\beta}(z,\bar
z)\xi^{\alpha}\xi^{\bar\beta}
\end{equation}
We will refer to the expression in the exponent as to the "action"
depending on the commuting complex coordinates $(z^{\alpha},
z^{\bar\alpha})$ of the manifold $\sf{M}$ and on the anticommuting
variables $\left(\xi^{\alpha}, \xi^{\bar\alpha}\right)$.

Let us introduce  a first-order differential operator $D$ defined
by the formula
\begin{equation}\label{D}
D=\xi^{\alpha}\partial_{\alpha} +
\xi^{\bar\alpha}\partial_{\bar\alpha} -iw^{\alpha\bar\beta}(z,\bar
z) \left(\partial_{\alpha}H(z,\bar z)\partial_{\bar\beta}^{\xi} -
\partial_{\bar\beta} H(z,\bar z)\partial_{\alpha}^{\xi}\right)
\end{equation}
 Using explicit expressions given above one can verify that such a
differential operator annihilates the action $S$, i.e. $DS=0$.
Next step is to construct a function $\lambda(z,\bar z,\xi,
\bar\xi)$ on the extended space which is annihilated by a repeated
action of the operator $D$, i.e. $D^2\lambda=0$. When such a
function exists and   integral (\ref{integral}) converges it is
possible to deform  integral (\ref{intflat}) as follows,
\begin{equation}\label{intflatdeformed}
I\rightarrow I_t=\int\prod\limits_{\alpha,\bar\alpha=1}^{m}
dz^{\alpha}dz^{\bar\alpha}\prod\limits_{\beta,\bar\beta=1}^{m}
d\xi^{\beta}d\xi^{\bar\beta}\exp\left(S(z,\bar
z,\xi,\bar\xi)+tD\lambda(z,\bar z,\xi,\bar\xi)\right)
\end{equation}
where $t$ is an arbitrary parameter.
 Indeed, one can expand the
integrand in a series with respect to the parameter $t$ and use
the integration by parts together with the properties of the
differential operator $D$ ($DS=0, D^2\lambda=0$)  to verify that
the integral $I_t$ does not depend on the parameter $t$, i.e
$I_t=I$. Following  the general procedure described by
Zirnbauer\cite{Zirn} we make the following choice for the function
$\lambda (z,\bar z,\xi, \bar\xi)$:
\begin{equation}\label{lambda}
\lambda (z,\bar z,\xi, \bar\xi)=i\left(\partial_{\alpha}H(z,\bar
z)\xi^{\alpha}-\partial_{\bar\alpha}H(z,\bar
z)\xi^{\bar\alpha}\right)
\end{equation}
The action of the first-order differential operator $D$ on this
function gives
\begin{equation}\label{Dlambda}
D\lambda (z,\bar z,\xi,
\bar\xi)=-2\left(w^{\alpha\bar\beta}(z,\bar z)
\partial_{\alpha} H(z,\bar
z)\partial_{\bar\beta}H(z,\bar z)+i\partial_{\alpha\bar\beta}
H(z,\bar z)\xi^{\alpha}\xi^{\bar\beta}\right)
\end{equation}
A repeated  action of the operator $D$ on the function $\lambda$
gives the following expression:
\begin{equation}\label{D2lambda}
D^2\lambda (z,\bar z,\xi, \bar\xi)=-2\xi^{\alpha}\partial_{\alpha}
\left(w^{\mu\bar\nu}(z,\bar z)\partial_{\mu}H(z,\bar z)\right)+
2\xi^{\bar\alpha}\partial_{\bar\alpha}\left(w^{\mu\bar\nu}(z,\bar
z)\partial_{\bar\nu}H(z,\bar z)\right)
\end{equation}
It immediately follows that  the function $\lambda (z,\bar z,\xi,
\bar\xi)$ satisfies $D^2\lambda=0$ if and only if the Hamiltonian
$H(z,\bar z)$ satisfies the following conditions:
\begin{equation}\label{Hamiltonianconditions}
\partial_{\alpha}\left(w^{\mu\bar\nu}(z,\bar
z)\partial_{\mu}H(z,\bar z)\right)=0,\;\;\partial_{\bar\beta}
\left(w^{\mu\bar\nu}(z,\bar z)\partial_{\nu}H(z,\bar
z)\right)=0\;\;
\end{equation}
The property of the K\"ahler cosets that ensures localisation is
that one can always choose on them a sign-definite Riemann metric
(see, for example, Kobayashi and Nomizu \cite{kobayashi} and a
review article by Bordemann, Forger and Romer \cite{bordemann}).
When the metric is positive definite the numerical part of the
expression for $D\lambda $ proportional to
\begin{eqnarray}\label{dH2}
(dH)^2=w^{\alpha\bar\beta}(z,\bar z)\partial_{\alpha} H(z,\bar
z)\partial_{\bar\beta}H(z,\bar z)
\end{eqnarray}
is positive definite as well. It then follows that the limit
$t\rightarrow\infty$ localises the integral $I_t$ in
Eq.({\ref{intflatdeformed}) on the critical set $dH=0$. In turn, it
implies that the original integral $I$ is localized on the
critical set of the Hamiltonian $H$ as well. For the negative definite
case one can just set $t\to -\infty$ with the same result.

 As is seen from  the definition of the
Hamiltonian phase flow on a K\"{a}hler manifold, Eq.(\ref{phaseflow}),
conditions Eq.(\ref{Hamiltonianconditions}) mean
that the phase flow generated by the localizable  Hamiltonian is
holomorphic. Such phase flow preserves the K\"{a}hler metric, as
follows from definition (\ref{phaseflow}).

In particular, consider  a Hamiltonian on the K\"{a}hler manifold
$\sf{M}$ which can be represented as a linear combination of the
momentum maps $T_a(z,\bar z)$ satisfying the Eqs.(\ref{difmoment}),
\begin{equation}\label{Hsummaps}
H(z,\bar z)=\sum\limits_{a=1}^nc_aT_a(z,\bar z)
\end{equation}
Then the relation Eq.(\ref{difmoment1}) ensures that such
Hamiltonians conform to the conditions
(\ref{Hamiltonianconditions}). Hence they are localisable in the
sense of the Duistermaat-Heckman principle provided the integral
(\ref{integral}) converges and $(dH)^2$ defined by equation
(\ref{dH2}) is sign-definite.

\subsection{Localisable Hamiltonians on flag manifolds with
unitary transformation group} Let us consider first the case of
the flag manifold $\sf{G}/\sf{H}$, with $\sf{G}$ being a unitary
transformation group, $\sf{G}^{\dag}=\sf{G}^{-1}$. In any matrix
representation there exist projection matrices $\eta_j$ which
correspond to the elements $h_j$ of the Cartan subalgebra of the
Lie algebra $\bf{g}$. The projection matrices are defined by the
following set of equations
\begin{eqnarray}\label{projection}
\eta_j=\eta_j^{\dag}\;\;\;\qquad\eta_j^2=\eta_j\;\;\qquad\;\eta_j\hat{h}_k=\hat{h}_k\eta_j\;\;\qquad\forall
j,k=1,\ldots,r \nonumber\\
\eta_j\hat{e}_{-q}\eta_j=\hat{e}_{-q}\eta_j\;\;\;\;\eta_j\hat{e}_q\eta_j=\eta_j\hat{e}_q\qquad\qquad\quad
\end{eqnarray}
Here the hat stands for the matrix representation. When the Cartan
subalgebra elements are represented by $N\times N$ diagonal
matrices, the projection matrices are also diagonal, and any
$N\times N$ diagonal matrix is a linear combination of the
projection matrices and the unit $N\times N$ matrix, ${\bf 1}_N$.

The projection matrices were introduced by Bando, Kuratomo,
Maskawa and Uehara \cite{bando1} and Itoh, Kugo and Kunitomo
\cite{itoh} to construct explicit formulae for the K\"{a}hler
potentials. In particular it was found that in the case of a flag
manifold with a unitary transformation group ${\sf G}$ the most
general K\"{a}hler potential is a linear combination of scalar
functions (the fundamental K\"{a}hler potentials). Each fundamental
K\"{a}hler potential corresponds to a basis element of the Cartan
subalgebra of the Lie algebra of the group $\sf{G}$. The explicit
expression for the fundamental K\"{a}hler potential corresponding
to the basis element $h_i$ of the Cartan subalgebra is given by
\begin{equation}\label{funitarykpotential}
K_i(z,\bar z)=\ln\det\left(\eta_iu^{\dag}(z)
u(z)\eta_i+I-\eta_i\right),\;\;\;u(z)=\exp(z^q\hat{e}_q)
\end{equation}
This formula was used by Bar-Moshe and Marinov \cite{marinov1,
marinov2} to derive the equivariant momentum maps in terms of the
local complex coordinates on a K\"{a}hler manifold. The
expression obtained by Bar-Moshe and Marinov is
\begin{equation}\label{rhotrace}
T_a(z,\bar z)=-\mbox{Tr}\left(\rho(z,\bar
z)\hat{\tau_a}\right),\;\;\;\rho(z,\bar
z)=\sum\limits_{i=1}^rl_i\rho_i(z,\bar z)
\end{equation}
where $l_i$ are arbitrary constant coefficients and the matrices
$\rho_i(z,\bar z)$ are given by the following equation:
\begin{equation}\label{rhoexpression}
\rho_i(z,\bar z)=u(z)\eta_i \left(\eta_iu^{\dag}(z)
u(z)\eta_i+I-\eta_i\right)^{-1}\eta_iu^{\dag}(z),\qquad\;\;
u(z)\equiv\exp\left(z^q\hat{e}_q\right)\;\;\;
\end{equation}
The Hamiltonians on K\"{a}hler manifolds which can be represented
as a linear combination of the momentum maps are localisable.
For the corresponding integrals the Duistermaat-Heckman formula is
applicable as we have seen in the previous section. Using the
expressions Eq.(\ref{rhotrace}) for the momentum maps we thus
obtain the following formula for the localisable Hamiltonians:
\begin{equation}\label{localizableunitaryhamiltonians}
H(z,\bar z)=\sum\limits_{a=1}^{n}\sum\limits_{i=1}^{r}
v_{ai}\mbox{Tr}\left(\rho_i(z,\bar z)\hat{\tau}_a\right),
\end{equation}
where $v_{ai}$ are arbitrary constant coefficients. The matrices
$\rho_{i}(z,\bar z)$ are transformed under the action of the group
$\sf{G}$ on the manifold $\sf{M}$ as
\begin{equation}\label{rhotransformation}
\rho_i(z,\bar z)\rightarrow\rho_i(gz,\overline{gz})=g\rho_i(z,\bar
z)g^{\dag},\;\;\;g^{\dag}=g^{-1}
\end{equation}
The above transformation law for the matrices $\rho_i(z,\bar z)$
can be verified using the decomposition $gu(z)=u(gz)p(z,g)$
 which determines the action of the transformation
group $\sf{G}$ on its flag manifold (see Appendix C for explicit
calculation). Note that $\rho_i(0,0)=\eta_i$ and the projection
matrix $\rho_i(z,\bar z)$ at the point with the complex
coordinates $(z^{\alpha},z^{\bar\alpha})$ can be written as
\begin{equation}\label{rhoizbarz}
\rho_i(z,\bar z)=g(z,\bar z)\eta_ig^{-1}(z,\bar z).
\end{equation}
Here the group element $g(z,\bar z)\in{\sf{G}}$ represents a point
of the coset space ${\sf{G/H}}$ parameterized as described earlier
in this section.

A short inspection of expressions
Eq.(\ref{localizableunitaryhamiltonians}) and Eq.(\ref{rhoizbarz}) for
the particular case ${\sf{G}}={\sf{U}}(N)$ and $\sf{H}=\sf{T}$
makes it clear that the localizable Hamiltonian $H(z,\bar z)$  is
the same as that entering the Itzykson-Zuber integral, see Eq.
(\ref{izham}).

The construction described above can be taken over to the case of
pseudo-unitary transformation group without much modification.

\section{Duistermaat-Heckman localisation principle for
manifolds with pseudo-unitary transformation group}

We consider the flag manifolds $\sf{G}/\sf{T}$ where the
transformation group $\sf{G}$ is a pseudo-unitary Lie group with
its elements satisfying the condition (\ref{pseudounitarygroup}).
 Localisable Hamiltonians will be expressed as linear
combinations of momentum maps similar to the case of the unitary
transformation groups. The only new element is the presence of the
matrix $\lambda$. We begin with the formula for the fundamental
K\"{a}hler potentials (compare with Eq.
(\ref{funitarykpotential})):
\begin{equation}\label{kahlerpseudo}
K_i(z,\bar z)=\ln\det\left(\eta_iu^{\dag}(z)\lambda
u(z)\eta_i+{\bf 1}-\eta_i\right),\;\;\;u(z)=\exp(z^qe_q)
\end{equation}
Once the fundamental K\"{a}hler potentials are known it is
possible to find the equivariant momentum maps in terms of the
local complex coordinates on the given flag manifold.
We have found that the equivariant maps have the
same form as in the unitary case (see Eq.(\ref{rhotrace})), but
the projection matrices $\rho_i(z,\bar z)$ turned out to be
slightly different and are given by
\begin{equation}\label{projpseudo}
\rho_i(z,\bar z)=u(z)\eta_i\left(\eta_iu^{\dag}(z)\lambda
u(z)\eta_i+{\bf 1}-\eta_i\right)^{-1}\eta_iu^{\dag}(z)\lambda
\end{equation}
We note that in the pseudounitary case the element $u^{\dag}(z)$
in the above expression comes together with the matrix $\lambda$
in the same way as it does in the  formula Eq.(\ref{kahlerpseudo}) for
the fundamental K\"{a}hler potentials. Respectively, the
Hamiltonians given by the formula
Eq.(\ref{localizableunitaryhamiltonians}) with $\rho_i(z,\bar z)$
specified by  Eq.(\ref{kahlerpseudo}) are localisable
provided that the integral Eq.(\ref{integral}) converges. The
transformation law for the matrices $\rho_i(z,\bar z)$ turns out
to be of the same form as one for the unitary case:
\begin{equation}\label{pseudorhotransformation}
\rho_i(z,\bar z)\rightarrow\rho_i(gz,\overline{gz})=g\rho_i(z,\bar
z)g^{-1},\;\;\;g^{\dag}\lambda g=\lambda
\end{equation}
Then the same argumentation makes it evident
 that the natural counterpart of the Itzykson-Zuber Hamiltonian
is
\begin{equation}\label{NonUnitaryIZ}
H(g,X,Y)=\mbox{Tr}\left(g(z,\bar z)Xg^{-1}(z,\bar
z)Y\right),\;\;g(z,\bar z)\in {\sf{U}}(n_1,n_2)/{\sf{T}}
\end{equation}
This Hamiltonian is localisable provided that the diagonal
matrices $X, Y$ are chosen properly as to ensure that the integral
Eq.(\ref{IZpseudo}) converges.

 In what follows we apply the Duistermaat-Heckman theorem to
the integral Eq.(\ref{IZpseudo}) with the Hamiltonian given by the expression
Eq.(\ref{NonUnitaryIZ}) and evaluate it by using the method of
stationary phase. We conventionally refer to this procedure as to
the "semiclassical approximation", but the Duistermaat-Heckman
localisation principle ensures that such an approximation yields
the exact result.

The elements $g$ of the coset space $\sf{U(n_1,n_2)}/\sf{T}$
satisfy  the condition Eq.(\ref{pseudounitarygroup}) where
$\sf{G}=\sf{U(n_1,n_2)}$, and the matrix $\lambda$ is given by
\begin{equation}\label{lambda666}
\lambda=\left(
\begin{array}{cc}
  I_{n_1} & 0 \\
  0 & -I_{n_2}
\end{array}
\right)
\end{equation}
For our calculation we make use of the same complex
parameterization for the manifold  $\sf{G}=U(n_1,n_2)/\sf{T}$ as that
described in section 2. In the integral Eq.(\ref{IZpseudo}) the matrices
$X, Y$ are diagonal, with the complex variables $x_1,\ldots
,x_{n_1+n_2}$ and $y_1,\ldots ,y_{n_1+n_2}$ .

Let us find now the set of solutions for the equation $dH=0$
(saddle points). We get:
\begin{equation}\label{saddlepointsequation1}
dH=\mbox{Tr}\left(\delta
g\left[Y,g^{-1}Xg\right]\right)=0,\;\;\;\;\delta g\equiv g^{-1}dg
\end{equation}
Note that the expression $\mbox{Tr}(\textbf{x},\textbf{y})$, with
$\textbf{x}\in
\textbf{g}$, $\textbf{y}\in\textbf{g}$ is a quadratic form on the
Lie algebra $\textbf{g}$. Then from the condition
$\mbox{Tr}(\textbf{x},\textbf{y})=0$  follows that $\textbf{x}=0$
or $\textbf{y}=0$. We see that  the equation
(\ref{saddlepointsequation1}) is equivalent to the condition of
the Lie algebra elements $Y$ and $g^{-1}Xg$ commuting with each
other, i.e.
\begin{equation}\label{saddlepointsequation2}
\left[Y,g^{-1}Xg\right]=0
\end{equation}

Once the matrices $X,Y$ are diagonal all the saddle points on the
homogeneous manifold $\sf{U}(n_1,n_2)/\sf{T}$ belong to the
permutation group $\sf{S_{n_1+n_2}}$ (which is the permutation
group of the diagonal entries of $X$ and $Y$). Denote these saddle
points by $P$. The saddle points $P$ should belong to the coset
space $\sf{U}(n_1,n_2)/\sf{H}$  so they are elements of the
pseudo-unitary group $\sf{U}(n_1,n_2)$ and  satisfy the following
constraint:
\begin{equation}\label{saddlepoints}
P^{\dag}\lambda P=\lambda,\;\;\;P\in {\sf{S}_{n_1+n_2}}
\end{equation}
It is clear that only permutation matrices that do not mix the
elements $+1$ and $-1$ of the matrix $\lambda$ could satisfy the
condition Eq.(\ref{saddlepoints}). Thus the relevant saddle points
are all elements of the permutation group $\sf{S_{n_1}}\times
\sf{S_{n_2}}$, where $\sf{S_{n_1}}$ stands for the permutation
group of the top $n_1$ diagonal elements of the matrix $\lambda$,
and $\sf{S_{n_2}}$ is the permutation group of the rest $n_2$
diagonal entries of $\lambda$. Let us recall that in the case of
\textit{unitary} coset space the relevant saddle points were
 all possible elements of $\sf{S_{n_1+n_2}}$. We conclude that the condition
ensuring the saddle points to be elements of the pseudo-unitary
group reduces the initial symmetry group $\sf{S_{n_1+n_2}}$ of
saddle points down to the symmetry group of the lowest order,
$\sf{S_{n_1}\times S_{n_2}}$ .

In order to evaluate the integral Eq.(\ref{IZpseudo}) by the stationary
phase method we rewrite it as a sum over contributions
from neighborhoods of the saddle points
\begin{equation}\label{integralsum}
I\equiv\int\limits_{g\in
{\sf{U(n_1,n_2)}}/\sf{T}}d\mu(g)\exp\left(iH(g,X,Y)\right)=
\sum\limits_{P\in \sf{S_{n_1}\times S_{n_2}}}\int
d\mu(g_p)\exp\left(iH(g_P,X,Y)\right)
\end{equation}
where $g_P=g\cdot P$. Each integral in the sum above corresponds
to one saddle point $P$ and should be taken over a neighborhood of
the point $P$ on the manifold ${\sf{U}(n_1,n_2)}/{\sf{T}}$.
Consider the shift  $g_P=g\cdot P$ as a change of coordinates.
Then it suffices to take the shift element $g$ to be close to the
unit element of the group. This allows one to explore the
neighborhood of the chosen saddle point solution in line with the
spirit of the Duistermaat-Heckman principle. Introducing the
complex parameterization $g=g(z,\bar z)$ we therefore take the
complex parameters of the shift element $g$ to be close to zero.
We note that the coset measure is invariant under the  group
shifts,
\begin{equation}\label{measure}
d\mu(g_P)=d\mu(g\cdot P)=d\mu(g)
\end{equation}
enabling us to use the same \textit{flat} measure at each integral
in the sum in the expression Eq.(\ref{integralsum}). Correspondingly, the
Hamiltonian $H(g_P,X,Y)$ is expanded
 in the vicinity of the relevant
saddle point $P$ separately at each of the integrals in the right
hand side of Eq.(\ref{integralsum}).

 For each $g\in \sf{U(n_1,n_2)}/\sf{T}$ represented by
 a point from the coset space we introduce
the local complex parameterization in the neighborhood of unity,
$g=I$. We have:
\begin{equation}\label{gsum}
g(0,0)=I,\;\;\; g(z,\bar z)=I+g^{(1)}(z,\bar z)+g^{(2)}(z,\bar
z)+O\left(|z|^3\right)
\end{equation}
(Here $g^{(1)}(z,\bar z)=O(|z|)$ and $g^{(2)}(z,\bar
z)=O(|z|^{2})$). The  condition of pseudounitarity of the
transformation group underlying the given coset space yields:
\begin{equation}\label{gminussum}
g^{-1}(z,\bar z)=I+\lambda\left(g^{(1)}(z,\bar
z))\right)^{\dag}\lambda+\lambda g^{(2)}(z,\bar
z)\lambda+O\left(|z|^3\right)
\end{equation}

Then it is straightforward to find the Lie algebra element
$g^{(1)}(z,\bar z)$ up to the lowest order in $z, \bar z$ using
the Cartan-Weyl basis of the Lie algebra $\textbf{g}$:
\begin{eqnarray}\label{g1decomposition}
g^{(1)}(z,\bar z)=Z+N+Q\qquad\qquad\qquad\nonumber\\
Z=z^{q}e_q,\;\;\; N={\it N}^{q}(z,\bar z)e_{-q},\;\;\; Q={\it
Q}^j(z,\bar z)h_j
\end{eqnarray}
The Lie algebra elements $N$ and $Q$ turn out to be of the first
order in $|z|$. Insert Eqs.(\ref{gsum}), (\ref{gminussum})
with $g^{(1)}(z,\bar z)$ given by Eq.(\ref{g1decomposition})
to the pseudo-unitary condition Eq.(\ref{pseudounitarygroup}).
We neglect the terms of the second
order in $|z|$ and find the Lie algebra  elements $N$ and $Q$:
\begin{equation}\label{N}
 N=-\lambda Z^{\dag}\lambda,\;\;\; Q=0
 \end{equation}
We see that the first-order Lie algebra elements in terms of
 the local complex coordinates are given by:
\begin{eqnarray}\label{g1Z}
g^{(1)}(z,\bar z)=Z-\lambda
Z^{\dag}\lambda\equiv Z-Z_{\lambda}^{\dag}\\
Z=\sum\limits_{q=1}^{\frac{n-r}{2}}z^{q}e_q=\sum\limits_{1\leq
i<j\leq n_1+n_2}z_{ij}e_{(ij)}
\end{eqnarray}
where the standard basis elements $e_{(ij)}$ of the Lie algebra
are defined in the matrix representation by the formula:
\begin{equation}\label{standardbasis}
\left\{e_{(ij)}\right\}_{kl}=\delta_{ik}\delta_{jl}
\end{equation}
Now we should expand the Hamiltonian
\begin{equation}\label{hamiltonian}
H\left(g_{P}(z,\bar z)\right)\equiv\mbox{tr}\left(Xg_P(z,\bar
z)Yg_{P}^{-1}(z,\bar z)\right)
\end{equation}
in the vicinity of the saddle point $P$. For this purpose we
insert Eqs.(\ref{gsum}), (\ref{gminussum}) into the
Hamiltonian and obtain
\begin{eqnarray}\label{hamiltonianexpansion}
H\left(g_{P}(z,\bar z)\right)=H^{(0)}+H^{(1)}(z,\bar z)
+H^{(2)}(z,\bar z)+O\left(|z|^3\right)\;\;\;\;\qquad\qquad\nonumber\\
H^{(0)}=\mbox{tr}\left(XY_{P}\right)\qquad\qquad\qquad\qquad\qquad\qquad\qquad\nonumber\\
H^{(1)}(z,\bar
z)=\mbox{tr}\left(XY_{P}\left(\left(g_{\lambda}^{(1)}(z,\bar
z)\right)^{\dag}+g^{(1)}(z,\bar z)\right)\right)
\qquad\qquad\qquad\nonumber\\
H^{(2)}(z,\bar z)=\mbox{tr}\left(Xg^{(1)}(z,\bar
z)Y_{P}\left(g_{\lambda}^{(1)}(z,\bar z)\right)^{\dag}\right)
+\mbox{tr}\left(XY_{P}\left(\left(g_{\lambda}^{(2)}(z,\bar
z)\right)^{\dag}+g^{(2)}(z,\bar z)\right)\right)
\end{eqnarray}
where we have used the following notations:
\begin{equation}\label{gnot}
(g_{\lambda}^{(1)})^{\dag}\equiv\lambda
(g^{(1)})^{\dag}\lambda,\qquad
(g_{\lambda}^{(2)})^{\dag}\equiv\lambda
(g^{(2)})^{\dag}\lambda,\qquad Y_P\equiv PYP^{-1}
\end{equation}
The explicit forms for $H^{(1)}$ and $H^{(2)}$ in terms of the
local complex coordinates can be found from the pseudo-unitary
condition (\ref{pseudounitarygroup}). We find that $H^{(1)}=0$ and
$H^{(2)}$ can be written as
\begin{equation}\label{H2}
H^{(2)}(z,\bar z)=\mbox{tr}\left(Xg^{(1)}(z,\bar
z)Y_{P}\left(g_{\lambda}^{(1)}(z,\bar z)\right)^{\dag}\right)
-\mbox{tr}\left(XY_{P}\left(\left(g_{\lambda}^{(1)}(z,\bar
z)\right)^{\dag}g^{(1)}(z,\bar z)\right)\right)
\end{equation}
We further observe that taking into account terms of first and
second order with respect to $|z|$ results in an expression for the
Hamiltonian $H(g_{P}(z,\bar z))$ which depends only on the element
$g^{(1)}(z,\bar z)$.  Once the expression for $g^{(1)}(z,\bar z)$
is found (Eq.(\ref{g1Z})) we obtain from
Eqs.(\ref{hamiltonianexpansion}) and (\ref{H2}):
\begin{equation}\label{hamiltonianZ}
H(g_{P}(z,\bar z))=\mbox{tr}\left(XY_{P}\right)+
\mbox{tr}\left(Z^{\dag}_{\lambda}X\left[ZY_{P}\right]+
ZX\left[Z^{\dag}_{\lambda}Y_{P}\right]\right)+O\left(|z|^3\right)
\end{equation}
Inserting  $X=\sum\limits_{i=1}^{n_1+n_2}x^{i}e_{(ii)}$,
$Y_{P}=\sum\limits_{i=1}^{n_1+n_2}y_{P}^{i}e_{(ii)}$ and
$Z=\sum\limits_{1\leq i<j\leq n_1+n_2}z_{ij}e_{(ij)}$ to the above
equation and calculating the traces yields the Hamiltonian
$H(g_{P}(z,\bar z))$ up to the second order terms in $|z|$:
\begin{eqnarray}\label{Hzbarz}
H(g_{P}(z,\bar z))= \sum\limits_{i=1}^{n_1+n_2}x^iy^{i}_{P}-
\sum\limits_{1\leq i<j\leq n_1}(x^i-x^j)(y^i_P-y^j_P)z_{ij}\bar
z_{ij}\qquad\qquad\qquad\qquad\nonumber\\+ \sum\limits_{1\leq
i<n_1<j\leq n_1+n_2}(x^i-x^j)(y^i_P-y^j_P)z_{ij}\bar z_{ij}-
\sum\limits_{n_1+1\leq i<j\leq
n_1+n_2}(x^i-x^j)(y^i_P-y^j_P)z_{ij}\bar
z_{ij}+O(\left|z|^3\right)
\end{eqnarray}
We use the above expansion of the Hamiltonian $H(g_{P}(z,\bar z))$
in the vicinity of the saddle point $P$ to obtain the
semiclassical approximation for the integral Eq.(\ref{IZpseudo}):
\begin{eqnarray}\label{haintgauss}
I=\sum\limits_{P\in S_{n_1}\times
S_{n_2}}\exp\left(i\sum\limits_{j=1}^{n_1+n_2}x^jy^j_{P}\right)
\int\prod\limits_{1\leq i<j\leq n_1}\frac{dz_{ij}d\bar
z_{ij}}{(-2\pi i)}\; \exp\left(-i\sum\limits_{1\leq i<j\leq
n_1}(x^i-x^j)(y^i_P-y^j_P)z_{ij}\bar z_{ij}\right)\nonumber\\
\int\prod\limits_{n_1+1\leq i<j\leq n_1+n_2}\frac{dz_{ij}d\bar
z_{ij}}{(-2\pi i)}\; \exp\left(-i\sum\limits_{n_1+1\leq i<j\leq
n_1+n_2}(x^i-x^j)(y^i_P-y^j_P)z_{ij}\bar
z_{ij}\right)\qquad\qquad\;\;\;\;
\nonumber\\
\int\prod\limits_{1\leq i\leq n_1<j\leq n_1+n_2}\frac{dz_{ij}d\bar
z_{ij}}{(-2\pi i)}\; \exp\left(i\sum\limits_{1\leq i\leq n_1<j\leq
n_1+n_2}(x^i-x^j)(y^i_P-y^j_P)z_{ij}\bar
z_{ij}\right)\qquad\qquad\;\;\;\;
\end{eqnarray}
Calculating the Gaussian integrals and noting that for the
Vandermonde determinantal factors $\Delta(Y)=\prod_{i<j}(y_i-y_j)$
one has $\bigtriangleup\left(Y_P\right)=(-)^{\epsilon}
\bigtriangleup\left(Y\right)$, with $\epsilon$ being equal to
unity (zero) for even (odd) permutations, respectively, after
straightforward manipulations we obtain the final formula:
\begin{equation}\label{integralresult}
I=(-1)^{n_1n_2}\frac{\det\left(\exp(ix^ky^l)\right)|_{1\leq
k,l\leq n_1} \det\left(\exp(ix^ky^l)\right)|_{n_1+1\leq k,l\leq
n_1+n_2}}{\bigtriangleup\left(X\right)\bigtriangleup\left(Y\right)}
\end{equation}

\section{Correlation function of characteristic polynomials:
 general formalism}

In the present section we derive an integral representation for
the correlation function of characteristic polynomials of the GUE
matrices suitable for further investigation in the limit of large
matrix dimensions. Our method is a refined version of that
introduced in [I].

Let $\hat{H}$ be $N\times N$ random Hermitian matrix $\hat{H}=
\hat{H}^{\dagger}$ which is
characterized by the standard (GUE) joint probability density:
\begin{equation} \label{GUE}
{\cal P}(\hat{H})=C_N\exp{-\frac{N}{2}\mbox{Tr}\hat{H}^2},\quad
C_N=(2\pi)^{-\frac{N(N+1)}{2}}N^{N^2/2}
\end{equation}
with respect to the measure
$d\hat{H}=\prod_{i=1}^N dH_{ii}\prod_{i<j}dH_{ij}d\overline{H}_{ij}$,
where as usual
$dzd\overline{z}\equiv 2d\mbox{Re}z d\mbox{Im} z$.

Regularizing the characteristic polynomial
$Z_N(\mu)=\det\left(\mu {\bf 1}_N-\hat{H}\right)$
by considering the spectral parameter $\mu$ such that
$\mbox{Im}\mu\ne 0$
we are interested in calculating the following generating function:
\begin{equation}
{\cal K}_{N}(\hat{\mu}_B,\hat{\mu}_F)=
\left\langle
\frac{\prod_{k=1}^{n_F} Z_N(\mu^{(k)}_{1F})Z_N(\mu^{(k)}_{2F})}
{\prod_{l=1}^{n_B}Z_N(\mu^{(l)}_{1B})Z_N(\mu^{(l)}_{2B})}
\right\rangle_{GUE}
\end{equation}
where we denote by $\left\langle ...\right\rangle$ the expectation value
with respect to the distribution Eq.(\ref{GUE}) and
\[
\hat{\mu}_B=\mbox{diag}(\mu^{(1)}_{1B},...,\mu^{(n_B)}_{1B},
\mu^{(1)}_{2B},...,\mu^{(n_B)}_{2B})\,, \,\,
\hat{\mu}_F=\mbox{diag}(\mu^{(1)}_{1F},...,\mu^{(n_F)}_{1F},
\mu^{(1)}_{2F},...,\mu^{(n_F)}_{2F})
\]
 and $\mbox{Im}(\mu^{(l)}_{1B},-\mu^{(l)}_{2B})>0$.

The generating function is obviously an analytic one
with respect to  the complex variables $(\mu^{(k)}_{1F},\mu^{(k)}_{2F})$.
 It  turns out to be technically convenient to change:
$\mu^{(k)}_{1F}\to -i\mu^{(k)}_{1F}\,\,,\,\,\mu^{(k)}_{2F}
\to -i\mu^{(k)}_{2F} $
when performing the ensemble averaging, and restore the original generating function
by a simple analytical continuation.

To calculate the average we first use the standard
"supersymmetrisation" procedure and  represent each of the
characteristic polynomials in the denominantor as the Gaussian
integrals:
\begin{eqnarray} \label{Gau}
[Z_N(\mu)]^{-1}
&=&\frac{1}{(4\pi i)^{N}}\int
d^2{\bf S}
\exp\left\{\pm\frac{i}{2}\left(\mu {\bf S}^{\dagger}
{\bf S}-{\bf S}^{\dagger}\hat{H}
{\bf S}\right)\right\}\\
\nonumber &=&
\frac{1}{(4\pi i)^{N}}\int
d^2{\bf S}
\exp\left\{\pm\frac{i}{2}\left(\mu {\bf S}^{\dagger}
{\bf S}-\mbox{Tr}\left[\hat{H}
{\bf S}\otimes{\bf S}^{\dagger}\right]\right)\right\}
\end{eqnarray}
where we introduced a complex $N-$dimensional
vector ${\bf S}=(s_{1},...,s_{N})^T$ (here $T$ stands for the vector
transposition) so that
$d^2{\bf S}=\prod_{i=1}^N ds_{i}d\overline{s}_{i}$.
The sign $\pm$ in the exponential
is coordinated with the sign of Im$\mu$ as to ensure convergency
of the integral.

For the characteristic polynomials
in the numerator we use the representation in terms of the
 Gaussian integrals over anticommuting
(Grassmannian) $N-$component vectors ${\bf \chi},{\bf
\chi}^{\dagger}$, see e.g. \cite{VZ}.
Taking the product of all the integrals,
the generating function can be written down in the
following form:
\begin{eqnarray}\nonumber
{\cal K}_{N}(\hat{\mu}_b,\hat{\mu}_f)&\propto&
\int \prod_k^{n_F}d\chi_{k,1} d\chi_{k,1}^{\dagger}\int
\prod_k^{n_F} d\chi_{k,2} d\chi_{k,2}^{\dagger}
\exp\left\{\frac{1}{2}\sum_k^{n_F}\left(\mu^{(k)}_{1F}
\chi_{k,1}^{\dagger}\chi_{k,1}+
\mu^{(k)}_{2F}\chi_{k,2}^{\dagger}\chi_{k,2}\right)\right\}
\\ &\times&
\int d^2{\bf S}_{l,1} \int d^2{\bf S}_{l,2}
\exp\left\{\frac{i}{2}\sum_{l=1}^{n_B}\left[
\mu^{(l)}_{1B} {\bf S}^{\dagger}_{l,1}
{\bf S}_{l,1}-\mu^{(l)}_{2B}{\bf S}^{\dagger}_{l,2}
{\bf S}_{l,2}\right]\right\}
\\ \nonumber &\times&
\left\langle \exp{\left[-\frac{i}{2}\mbox{Tr}\hat{H}
\left\{\sum_{l=1}^{n_B}\left(
 {\bf S}_{l,1}\otimes
{\bf S}^{\dagger}_{l,1}-{\bf S}_{l,2}
\otimes {\bf S}^{\dagger}_{l,2}\right)+
\sum_k^{n_F}
\left(\chi_{k,1}\otimes\chi_{k,1}^{\dagger}+
\chi_{k,2}\otimes\chi_{k,2}^{\dagger}\right)
\right\}\right]}\right\rangle_{GUE}
\end{eqnarray}
The ensemble average is then easy to perform
via the identity:
\[
\left\langle e^{-\frac{i}{2}\mbox{Tr}\left[\hat{H}\hat{A}
\right]}\right\rangle_{GUE}\propto
e^{-\frac{1}{8N}\mbox{Tr}\left[\hat{A}^2
\right]}
\]
and
after straightforward manipulations one arrives at:

\begin{eqnarray}\nonumber
{\cal K}_{N}(\hat{\mu}_b,\hat{\mu}_f)&\propto&
\int \prod_k^{n_F}d\chi_{k,1} d\chi_{k,1}^{\dagger}\int
\prod_k^{n_F} d\chi_{k,2} d\chi_{k,2}^{\dagger}
\exp\left\{\frac{1}{2}\sum_k^{n_F}\left(\mu^{(k)}_{1F}
\chi_{k,1}^{\dagger}\chi_{k,1}+
\mu^{(k)}_{2F}\chi_{k,2}^{\dagger}\chi_{k,2}\right)
+\frac{1}{8N}\mbox{Tr}\left(\hat{Q}_F^2\right)\right\}
\\ &\times&
\int d^2{\bf S}_{l,1} \int d^2{\bf S}_{l,2}
\exp\left\{\frac{i}{2}\sum_{l=1}^{n_B}\left[
\mu^{(l)}_{1B} {\bf S}^{\dagger}_{l,1}
{\bf S}_{l,1}-\mu^{(l)}_{2B}{\bf S}^{\dagger}_{l,2}
{\bf S}_{l,2}\right]
-\frac{1}{8N}\mbox{Tr}\left(\tilde{Q}_B\hat{L}\tilde{Q}_B\hat{L}
\right)\right\}\\ \nonumber &\times&
\exp\left\{-\frac{1}{4N}
\sum_{k=1}^{n_F}\sum_{p=1}^2\chi_{k,p}^{\dagger}
\left(\sum_{l=1}^{n_B}{\bf S}_{l,1}\otimes{\bf S}^{\dagger}_{l,1}-
{\bf S}_{l,2}\otimes{\bf S}^{\dagger}_{l,2}\right)\chi_{k,p}\right\}
\end{eqnarray}
where we introduced the following $2n_F\times 2n_F$ Hermitian matrices $\hat{Q}_F$
and $2n_B\times 2n_B$ Hermitian matrices $\tilde{Q}_B$:
\[
\hat{Q}_F=\left(\begin{array}{cc}\hat{Q}^{(11)}_F&\hat{Q}^{(12)}_F
\\ \hat{Q}^{(21)}_F& \hat{Q}^{(22)}_F
\end{array}\right)\quad,\quad
\hat{Q}_B=\left(\begin{array}{cc}\tilde{Q}^{(11)}_B&\tilde{Q}^{(12)}_B
\\ \tilde{Q}^{(21)}_B& \tilde{Q}^{(22)}_B
\end{array}\right)
\]
with entries
\[
 \left[\tilde{Q}^{(p_1p_2)}_F\right]_{l_1l_2}=
{\bf S}^{\dagger}_{l_1,p_1}{\bf S}_{l_2,p_2}
\quad,\quad
\left[\hat{Q}^{(q_1q_2)}_F\right]_{k_1k_2}=
{\bf \chi}^{\dagger}_{k_1,q_1}{\bf \chi}_{k_2,q_2}
\]
and used the notation
$\hat{L}=\mbox{diag}({\bf 1}_{n_B},-{\bf 1}_{n_B})$

Now we employ the Hubbard-Stratonovich identity:
\begin{equation}
\exp{\left[\frac{1}{8N}\mbox{Tr}\hat{Q}_F^2\right]}\propto\int d\tilde{Q}_F
\exp{\left[-\frac{N}{2}\mbox{Tr}\tilde{Q}_F^2-\frac{1}{2}
(\chi_{k,1}^{\dagger},\chi_{k,2}^{\dagger})
\tilde{Q}^T_F\left(\begin{array}{c}\chi_{k,1}\\ \chi_{k,2}\end{array}\right)\right]}
\end{equation}
where the integration goes over the manifold of $2n_F\times 2n_F$
Hermitian matrices
$\tilde{Q}_F=\left(\begin{array}{cc}\hat{q}^{(11)}_F&\hat{q}^{(12)}_F
\\ \hat{q}^{(21)}_F& \hat{q}^{(22)}_F\end{array}\right)$
with the symmetry structure inherited from that of $\hat{Q}_F$
and we introduced the shorthand notation:
\[
(\chi_{k,1}^{\dagger},\chi_{k,2}^{\dagger})\equiv
\left(\chi_{1,1}^{\dagger},...,
\chi_{n_F,1}^{\dagger},\chi_{1,2}^{\dagger},...,\chi_{n_F,2}^{\dagger}\right)
\]

Exploiting the above identity allows one to
perform the gaussian Grassmannian integral explicitly and
to bring the expression to the form:
\begin{eqnarray}\label{prom}
\nonumber
{\cal K}_{N}(\hat{\mu}_b,\hat{\mu}_f)&\propto&
\int d\tilde{Q}_F
e^{-\frac{N}{2}\mbox{Tr}\tilde{Q}_F^2}
\int \prod_{l=1}^{n_B}d^2{\bf S}_{l,1}d^2{\bf S}_{l,2}
\exp\left\{\frac{i}{2}\sum_{l=1}^{n_B}\left(
\mu^{(l)}_{1B} {\bf S}^{\dagger}_{l,1}
{\bf S}_{l,1}-\mu^{(l)}_{2B}{\bf S}^{\dagger}_{l,2}
{\bf S}_{l,2}\right)\right\}\\
&\times& e^{-\frac{1}{8N}\mbox{Tr}\left(\tilde{Q}_B\hat{L}\tilde{Q}_B\hat{L}
\right)}
\mbox{det}\hat{A}_B
\end{eqnarray}
where
\begin{equation}
\hat{A}_B=\left(\begin{array}{cccc}a_{1,1}{\bf 1}_N-
\hat{B} &a_{1,2}{\bf 1}_N&...&a_{1,2n_F}{\bf 1}_N\\
a_{2,1}{\bf 1}_N & a_{2,2}{\bf 1}_N-
\hat{B} &...&a_{1,2n_F}{\bf 1}_N\\
... & ... & ... & ...\\
a_{1,2n_F}{\bf 1}_N &
a_{2,2n_F}{\bf 1}_N & ... &a_{2n_F,2n_F}{\bf 1}_N-\hat{B}
\end{array}\right)
\end{equation}
We have used notations $a_{k_1,k_2}=\mu_F^{(k)}\delta_{k_1,k_2}-
(q_F)_{k_1k_2}$, for $k=1,2,...,n_F$ and introduced the $N\times
N$ matrix $\hat{B}=\frac{1}{2N}\sum_{l=1}^{n_B} \left[{\bf
S}_{l,1}\otimes{\bf S}^{\dagger}_{l,1}- {\bf S}_{l,2}\otimes{\bf
S}^{\dagger}_{l,2}\right]$.

To bring the determinant of the matrix $\hat{A}_B$
 in the above expession
to the form suitable for further manipulations we
consider the case $N\ge 2n_B$ and further introduce the
$N\times N$ matrix
\[
\hat{M}_N=\left(\hat{M},{\bf e_1},{\bf e_2},...,{\bf
e_{N-2n_B}} \right)
\]
where the N-component orthonormal vectors ${\bf e_1},{\bf e_2},...,{\bf
e_{N-2n_B}}$ are chosen to form a basis of
 the orthogonal complement to the linear span
of vectors ${\bf S}_{1,1},...,{\bf S}_{n_B,1},
{\bf S}_{1,2},...,{\bf S}_{n_B,2}$
(without restricting generality we can consider the latter
vectors to be linear independent).
Correspondingly, the matrix $\hat{M}$ is chosen to have
$2n_B$ columns which are just the vectors ${\bf S}_{l,1}$ and ${\bf
S}_{l,2}$. Simple calculation shows that
\begin{equation}
\hat{M}^{\dagger}\hat{M}=\tilde{Q}_B\quad\mbox{and} \quad
\hat{M}^{\dagger}\hat{B}\hat{M}=\frac{1}{2N}\tilde{Q}_B\hat{L}\tilde{Q}_B
\end{equation}
Then we can write:
\begin{eqnarray}
&& \mbox{det}\left(\begin{array}{cccc}\hat{M}_N^{\dagger}&0 & &\\
0  & \hat{M}_N^{\dagger} & & \\ ... &...& ... & 0 \\
&  & 0 & \hat{M}_N^{\dagger}\end{array}\right)\times
\mbox{det}( \hat{A}_B)\times
\mbox{det}\left(\begin{array}{cccc}\hat{M}_N& 0& & \\
0 & \hat{M}_N & &  \\ ... &...& ...& 0 \\
&  &0 & \hat{M}_N\end{array}\right)
\\ \nonumber
&=&\mbox{det}\left[a_{k_1,k_2}\left(\begin{array}{cc}
\tilde{Q}_B &\\&{\bf 1}_{N-2n_B}\end{array}\right)
-\delta_{k_1,k_2}\frac{1}{2N}\left(\begin{array}{cc}
\tilde{Q}_B\hat{L}\tilde{Q}_B &\\&{\bf 1}_{N-2n_B}\end{array}\right)
\right]\\ \nonumber &=&
\mbox{det}^{N-2n_B}
\left(\begin{array}{cccc}a_{1,1} &a_{1,2}&...&a_{1,2n_F}\\
a_{2,1} & a_{2,2} &...&a_{1,2n_F}\\
... & ... & ... & ...\\a_{1,2n_F} &
a_{2,2n_F} & ... &a_{2n_F,2n_F}
\end{array}\right)
\mbox{det}\left(a_{k_1,k_2}\tilde{Q}_B-
\delta_{k1,k_2}\frac{1}{2N}
\tilde{Q}_B\hat{L}\tilde{Q}_B\right)
\end{eqnarray}
Now it is easy to see that the determinant factor
$\mbox{det}{\hat{A}_B}$ is given by
\[
\mbox{det}\hat{A}_B=
\mbox{det}\left[\hat{\mu}_F-\tilde{Q}_F\right]^{N-2n_B}
\prod_{k=1}^{2n_F}\det{\left[q^{\mu}_{kF}{\bf 1}_{2n_b}-
\frac{1}{2N}\tilde{Q}_B\hat{L}
\right]}
\]
where we introduced the notation $q^{(\mu)}_{k,F}$
for (real) eigenvalues of the (Hermitian)
matrix $\tilde{Q}^{(\mu)}_F=\hat{\mu}_F-\tilde{Q}_F$.

Next step is to deal with the integrals
over ${\bf  S}_{l,1},{\bf  S}_{l,2}$.
For this we observe that the integrand
depends on those variables only via the matrix  $\tilde{Q}_B$
and employ the following\\
{\bf Theorem I}

{\sf
Consider a function $ F({\bf S_1},...,{\bf S_m})$
of $N$-component complex vectors ${\bf S}_l\,\, 1\le l\le m$ such that
\begin{equation}\label{conv}
\int_{C^N}d^2{\bf S}_1...\int_{C^N}d^2{\bf S}_m
|F({\bf S_1},...,{\bf S_m})|<\infty
\end{equation}
Suppose further that the function $F$
depends only on $m^2$ scalar products ${\bf S}^{\dagger}_{l_1}
{\bf S}_{l_2}\,\, 1\le l_1,l_2\le m$ so that it can be rewritten as
a function ${\cal F}(\tilde{Q}_m)$ of
$m\times m$ Hermitian matrix $\tilde{Q}_m$:
\[
\tilde{Q}_m=\left(\begin{array}{ccccc}
{\bf S}^{\dagger}_{1}{\bf S}_{1}&{\bf S}^{\dagger}_{1}
{\bf S}_{2}& ... & ... &{\bf S}^{\dagger}_{1}
{\bf S}_{m}\\ {\bf S}^{\dagger}_{2}{\bf S}_{1}& ... & ... &...&{\bf S}^{\dagger}_{2}{\bf S}_{m}\\
... & ... &{\bf S}^{\dagger}_{l_1}{\bf S}_{l_2} &
 ...& ... \\ {\bf S}^{\dagger}_{m}{\bf S}_{1} &... & ...&...&
{\bf S}^{\dagger}_{m}{\bf S}_{m}\end{array}\right)
\]
Then for $N\ge m$
\begin{equation}\label{trans}
\int_{C^N}d^2{\bf S}_1...\int_{C^N}d^2{\bf S}_m
F({\bf S_1},...,{\bf S_m})={\cal C}_{N,m}
\int_{\hat{Q}_m>0} d\hat{Q}_m\left(\mbox{det}\hat{Q}_m\right)^{N-m}
{\cal F}(\hat{Q}_m)
\end{equation}
where
\[
{\cal C}_{N,m}=\frac{(2\pi)^{Nm-\frac{m(m-1)}{2}}}{\prod_{k=1}^m(N-k)!}
\]
and the integration in the right-hand side of Eq.(\ref{trans})
goes over the manifold
of Hermitian positive definite $m\times m$ matrices $\hat{Q}_m$.
}

In fact, the formula Eq.(\ref{trans}) was already implicitly used
in [I]. In that paper it was justified by heuristic arguments
employing the Fourier transform of the function ${\cal F}$ and
subsequent exploitation of a matrix integral close to that
considered by Ingham and Siegel\cite{IS}. A proof of the theorem
is given in the Appendix D of the present paper\footnote{After
completion of our work we learned that an equivalent formula was
used earlier by David, Duplantier and Guitter \cite{Dupl} in quite
a different context}.

In our particular case $m=2n_B$ and the role of $\tilde{Q}_m$ is
played by $\tilde{Q}_B$.
 The convergency condition Eq.(\ref{conv})
in Eq.(\ref{prom})
is ensured by imaginary parts of the spectral parameters $\mu$.
Applying the theorem, we get:
\begin{eqnarray}\label{prom1}
&& {\cal K}_{N}(\hat{\mu}_B,\hat{\mu}_F)\propto
\int d\tilde{Q}_F\left(\det{\tilde{Q}_F^{(\mu)}}\right)^{N-2n_B}
e^{-\frac{N}{2}\mbox{Tr}\tilde{Q}_F^2}\\ \nonumber &\times&
\int_{\tilde{Q}_B>0} d\tilde{Q}_B \det{\tilde{Q}_B}^{N-2n_B}
e^{-\frac{1}{8N}\mbox{Tr}\left(\tilde{Q}_B\hat{L}\right)^2+
\frac{i}{2}\mbox{Tr}\left[\hat{\mu}_B\tilde{Q}_B\hat{L}\right]}
\prod_{k=1}^{2n_F}
\det{\left[q^{(\mu)}_{kF}{\bf 1}_{2n_B}-\frac{1}{2N}\tilde{Q}_B\hat{L}
\right]}
\end{eqnarray}

Let us now replace the Hermitian matrix  $\tilde{Q}_F$
by $\tilde{Q}_F^{(\mu)}$ as the integration
manifold and further change: $ \tilde{Q}_B\to
{2N}\tilde{Q}_B$. Omitting both tilde and $(\mu)$ symbols henceforth
 we arrive at:
\begin{eqnarray}\label{prom3}
&& {\cal K}_{N}(\hat{\mu}_B,\hat{\mu}_F)\propto
\int dQ_F\left(\mbox{det}Q_F\right)^{N-2n_B}
e^{-\frac{N}{2}\mbox{Tr}[\hat{\mu}_F-Q_F]^2}\\ \nonumber &\times&
\int_{Q_B>0} dQ_B \left(\mbox{det}Q_B\right)^{N-2n_B}
e^{-\frac{N}{2}\mbox{Tr}\left(Q_B\hat{L}\right)^2+
i\,N \mbox{Tr}\left[\hat{\mu}_B\hat{Q}_B\hat{L}\right]}
\det{\left[Q_F\otimes{\bf 1}_{2n_B}-{\bf 1}_{2n_F}\otimes Q_B\hat{L}
\right]}
\end{eqnarray}

The integral representation Eq.(\ref{prom3}) is our main result
for the present section. It is valid for any parameters
$N,n_{B,F},\hat{\mu}_{B,F}$
 provided $N\ge 2n_B$. The form of the integrals is clearly
 suggestive of treating them by the saddle-point method in the limit
of large $N$.
The details of the procedure are described in the
following section.

\section{Large $N$ behavior of the correlation function}

To perform the saddle-point evaluation of the integrals we first
have to expose those degrees of freedom which are amenable to such a
treatment. It is immediately evident that for
the matrix $Q_F$ the relevant variables are real eigenvalues
$-\infty< q_{k} <\infty \,,1\le k\le 2n_F$.
Accordingly, we write $Q_F=U\hat{q}_FU^{\dagger}$, where $U\in
U(2n_F)$ is $2n_F\times 2n_F$ unitary matrix, and
$\hat{q}_F=\mbox{diag}(q_{1},...,q_{2n_F})$. The integration
measure in those variables is known to be written by
$dQ_F\propto \Delta^2\{\hat{q}_F\}
d\mu(U)d\hat{q}_F$, with $d\mu(U)$ being the corresponding
Haar's measure on the group $U(2n_F)$ and
 $\Delta\{\hat{q}_F\}=\prod_{k_1<k_2}\left(q_{k_1}-q_{k_2}\right)$
standing for the Vandermonde determinant (see e.g. \cite{Hua}).

The only term in the integrand of Eq.(\ref{prom3}) which depends on
the unitary matrix $U$ is obviously the exponential
$\exp{N\mbox{Tr}\left(\hat{\mu}_FU\hat{q}_FU^{\dagger}\right)}$.
We immediately see that the corresponding integral over the unitary
group is exactly that by Itzyson-Zuber-Harish-Chandra,
Eq.(\ref{IZ}). This yields:
\begin{eqnarray}\label{prom4}
&& {\cal K}_{N}(\hat{\mu}_B,\hat{\mu}_F)\propto
\frac{1}{\Delta\{\hat{\mu}_F\}}e^{-\frac{N}{2}\mbox{Tr}[\hat{\mu}_F]^2}
\int d\hat{q}_F\Delta\{\hat{q}\}\left(\mbox{det}\hat{q}_F\right)^{N-2n_B}
\mbox{det}\left[e^{N\mu^{k_1}_{F}q_{k_2}}\right]_{ k_1,k_2=1}^{2n_{F}}
\\ \nonumber &\times&
e^{-\frac{N}{2}\mbox{Tr}[\hat{q}_F]^2}
\int_{Q_B>0} dQ_B \left(\mbox{det}Q_B\right)^{N-2n_F}
e^{-\frac{N}{2}\mbox{Tr}\left(Q_B\hat{L}\right)^2+
i\,N \mbox{Tr}\left[\hat{\mu}_BQ_B\hat{L}\right]}
\prod_{k=1}^{2n_F}
\det{\left[q_{k}{\bf 1}_{2n_B}-Q_B\hat{L}
\right]}
\end{eqnarray}

It is of little utility, however, to introduce
eigenvalues/eigenvectors of $Q_B>0$ as the integration variables.
Rather, it is natural to treat
$Q^{(L)}_B=Q_B\hat{L}$ as a new matrix to integrate over.
Properties of these (non-Hermitian!) matrices
 are discussed at length in the Appendix B of [I], and
references therein.
The matrices satisfy $\left[Q_B^{(L)}\right]^{\dagger}=
\hat{L}Q_B^{(L)} \hat{L}$, have
all eigenvalues real and can be
diagonalized by a ({\it pseudo}unitary) similarity transformation:
$Q_B^{(L)}=\hat{T}\hat{p}_B\hat{T}^{-1}$, where
$\hat{p}_B=\mbox{diag}(\hat{p}_1,\hat{p}_2)$, and
$n_B\times n_B$ diagonal matrices $\hat{p}_1,\hat{p}_2$ satisfy:
$\hat{p}_1>0\,,\,\hat{p}_2<0$.
Pseudounitary matrices $\hat{T}$ satisfy:
$\hat{T}^{\dagger}\hat{L}\hat{T}=\hat{L}$ and form the group $U(n_B,n_B)$
("hyperbolic symmetry").

We again introduce
the diagonal entries $\hat{p}_{1}$ and $\hat{p}_{2}$  along with the matrices
$\hat{T}\in \frac{U(n_B,n_B)}{U(1)\times ...\times U(1)}$
as new integration variables.
The integration measure $dQ_B^{(L)}$
is given in new variables as \cite{VZ}:
\[
dQ_B^{(L)}
\propto d\hat{p}_1d\hat{p}_2 \prod_{l_1<l_2}^{n_B}
\left(p^{(l_1)}_{1}-p^{(l_2)}_{1}\right)^2
\left(p^{(l_1)}_{2}-p^{(l_2)}_{2}\right)^2
\prod_{l_1,l_2}\left(p^{(l_1)}_{1}-p^{(l_2)}_2\right)^2 d\mu(T)
\]
where the last factor is the invariant measure on the coset space
of $T-$matrices.

Again, the only term in the integrand of Eq.(\ref{prom3}) which depends on
the pseudounitary matrices $\hat{T}$ is obviously the exponential
$\exp{N\mbox{Tr}\left(\hat{L}\hat{\mu}_B\hat{T}\hat{p}_B\hat{T}^{-1}\right)}$.
We immediately see that the corresponding integral over the
non-compact ("hyperbolic") manifold of $T-$ matrices
 is exactly that addressed by us in Section I (cf. Eq.(\ref{IZpseudo})).
\begin{eqnarray}\label{coset}
&&I(\hat{\mu}_B,\hat{p}_1,\hat{p}_2)=\int d\mu(\hat{T})
\exp\left\{iN\mbox{Tr}\left(\begin{array}{cc}
\hat{\mu}_{1B}&\\ & \hat{\mu}_{2B}
\end{array}\right)\hat{T}
\left(\begin{array}{cc}\hat{p}_1&\\ & \hat{p}_2
\end{array}\right)\hat{T}^{-1}\right\}\\
\nonumber
&&\propto\displaystyle{
\frac{\mbox{det}\left[e^{iN\mu^{(l_1)}_{1B}p^{(l_2)}_1}
\right]_{l_1,l_2=1}^{n_{B}}
\mbox{det}\left[e^{-iN\mu^{(l_1)}_{2B}p^{(l_2)}_{2}}
\right]_{l_1,l_2=1}^{n_{B}}}
{\Delta\{\hat{\mu}_B\}\Delta\{\hat{p}_1\} \Delta\{\hat{p}_2\}
\prod_{l_1,l_2}\left(p^{(l_1)}_1-p^{(l_2)}_2\right)}}
\end{eqnarray}

As a final step we change $\hat{p}_2\to - \hat{p}_2$ and with the
integrand depending only on the eigenvalues, we arrive to the
following expression:
\begin{eqnarray}\label{k3}
\nonumber
&& {\cal K}_{N}(\hat{\mu}_B,\hat{\mu}_F)\propto
\frac{e^{-\frac{N}{2}\mbox{Tr}[\hat{\mu}_F]^2}}
{\Delta\{\hat{\mu}_F\}\Delta\{\hat{\mu}_B\}}
\int_{R^{(F)}}
 d\hat{q}_F\Delta\{\hat{q}_F\}\left(\mbox{det}\hat{q}_F\right)^{N-2n_B}
\mbox{det}\left[e^{N\mu^{k_1}_{F}q_{k_2}}\right]_{ k_1,k_2=1}^{2n_{F}}
\\ \nonumber &\times&
e^{-\frac{N}{2}\mbox{Tr}[\hat{q}_F]^2}
\int_{R^{(B)}_{+}}
 d\hat{p}_1\Delta\{\hat{p}_1\}\int_{R^{(B)}_{+}}d\hat{p}_2
\,\,\Delta\{\hat{p}_2\}
 \prod_{l_1,l_2=1}^{n_B}
\left(p^{(l_1)}_1+p^{(l_2)}_2\right)e^{-\frac{N}{2}\mbox{Tr}
\left(\hat{p}_1^2+\hat{p}_2^2\right)}
 \mbox{det}\left[\hat{p}_1\hat{p}_2\right]^{N-2n_B}
 \\ &\times&
\mbox{det}\left[e^{iN\mu^{(l_1)}_{1B}p^{(l_2)}_1}\right]_{l_1,l_2=1}^{2n_{B}}
\mbox{det}\left[e^{-iN\mu^{(l_1)}_{2B}p^{(l_2)}_2}\right]_{l_1,l_2=1}^{2n_{B}}
\prod_{k=1}^{2n_F}
\mbox{det}\left[q_{k}{\bf 1}_{n_B}-\hat{p}_1\right]
\mbox{det}\left[q_{k}{\bf 1}_{n_B}+\hat{p}_2\right]
\end{eqnarray}
where we denoted $R^{(F)}$ the integration domain: $-\infty<q_k
<\infty$, $k=1,...,2n_F$ and $R^{(B)}_{+}$ the domain $0\le
p^{(l)}<\infty$ for $l=1,2,...,n_B$. Taking into account presence
of the Vandermonde determinant antisymmetric in all $q$'s as well
as
 the symmetry of the rest of the integrand with respect to
$(2n_F)! $ permutations of entries of the matrix $\hat{q}_F=
\mbox{diag}(q_{1},...,q_{2n_F})$ we see that we can
effectively replace the determinant factor:
\[
\mbox{det}\left[e^{N\mu^{(k_1)}_{F}q_{k_2}}\right]_{ k_1,k_2=1}^{2n_{F}}
\longrightarrow (2n_F)! e^{N\sum_{k=1}^{2n_F}\mu^(k)_{F}q_{k}}
\]
and perform similar replacements for the other two determinants
in Eq.(\ref{k3}).

Summing up, we derived the following integral representation for
the correlation functions of the characteristic polynomials
\begin{eqnarray}\label{kk3}
\nonumber
&& {\cal K}_{N}(\hat{\mu}_B,-i\hat{\mu}_F)=
\left\langle
\frac{\prod_{k=1}^{n_F} Z_N(-i\mu^{(k)}_{1F})Z_N(-i\mu^{(k)}_{2F})}
{\prod_{l=1}^{n_B}Z_N(\mu^{(l)}_{1B})Z_N(\mu^{(l)}_{2B})}
\right\rangle_{GUE}
\\  &=& \tilde{C}
\frac{1}{\Delta\{\hat{\mu}_F\}\Delta\{\hat{\mu}_B\}}
\int_{R_F}
 d\hat{q}_F\Delta\{\hat{q}_F\}\left(\mbox{det}\hat{q}_F\right)^{N-2n_B}
e^{-\frac{N}{2}\sum_{k=1}^{2n_F}\left(\mu^{(k)}_{F}-q_{k}\right)^2}
\\ \nonumber &\times&
\int_{R^{(B)}_{+}}
 d\hat{p}_1\Delta\{\hat{p}_1\}\int_{R^{(B)}_{+}}d\hat{p}_2
\,\,\Delta\{\hat{p}_2\}
 \prod_{l_1,l_2=1}^{n_B}
\left(p^{(l_1)}_1+p^{(l_2)}_2\right)e^{-\frac{N}{2}\mbox{Tr}
\left(\hat{p}_1^2+\hat{p}_2^2\right)}
e^{iN\sum_{l=1}^{n_b}
\left[\mu^{(l)}_{1B}p^{(l)}_1-\mu^{(l)}_{2B}p^{(l)}_2\right]}
\\ \nonumber &\times&
 \mbox{det}\left[\hat{p}_1\hat{p}_2\right]^{N-2n_B}
\prod_{k=1}^{2n_F}
\mbox{det}\left[q_{k}{\bf 1}_{n_B}-\hat{p}_1\right]
\mbox{det}\left[q_{k}{\bf 1}_{n_B}+\hat{p}_2\right]
\end{eqnarray}
which is still exact for $N\ge 2n_B$ and valid for
arbitrary values of parameters
such that Im$\mu^{(l)}_{1B}>0$ and Im$\mu^{(l)}_{2B}<0$.

Before treating the integrals in the limit $N\to \infty$ by the
saddle-point method we can restore the normalisation constant
$\tilde{C}$ by comparing both sides of the equation in the limit
$N-$ fixed, $\mu_F^{(k)}\to \infty$, Im$\mu^{(l)}_{1B}\to \infty$,
Im$\mu^{(l)}_{2B}\to -\infty$. Obviously, in this limit the
presence of the random matrix $H\in GUE$ is immaterial and by its
very definition the correlation function
 tends to:
\begin{eqnarray}\label{lhs}
\nonumber
\left\langle
\frac{\prod_{k=1}^{n_F} Z_N(-i\mu^{(k)}_{1F})Z_N(-i\mu^{(k)}_{2F})}
{\prod_{l=1}^{n_B}Z_N(\mu^{(l)}_{1B})Z_N(\mu^{(l)}_{2B})}
\right\rangle_{GUE}\to (-1)^{Nn_F}
\frac{\left[\prod_{k=1}^{n_F} \mu^{(k)}_{1F}\mu^{(k)}_{2F}\right]^N}
{\left[\prod_{l=1}^{n_B}\mu^{(l)}_{1B}\mu^{(l)}_{2B}\right]^N}
\end{eqnarray}
In the right hand side close inspection shows that the integrals
over $q^{(k)}_F$ are dominated by vicinity of
$q^{(k)}_F=\mu^{(k)}_F$. They effectively decouple from the
integrals over $p_{1,2}^{(l)}$ and can be straightforwardly
calculated yielding exactly the factor:
\[
\Delta\{\hat{\mu}_F\}
\left(\frac{2\pi}{N}\right)^{n_F}\left[\prod_{k=1}^{n_F}
\mu^{(k)}_{1F} \mu^{(k)}_{2F}\right]^N
\]
On the other hand, performing the remaining integrals in the appropriate
limit amounts to evaluating the following expression:
\begin{eqnarray}
I=(-1)^{\frac{n_B(n_B+1)}{2}}\int_{R^{(B)}_{+}}
 d\hat{p}_1\int_{R^{(B)}_{+}}d\hat{p}_2
\,\,\Delta\{\hat{p}_1,-\hat{p}_2\}
e^{iN\sum_{l=1}^{n_b}
\left[\mu^{(l)}_{1B}p^{(l)}_1-\mu^{(l)}_{2B}p^{(l)}_2\right]}
 \mbox{det}\left[\hat{p}_1\hat{p}_2\right]^{N-2n_B}
\end{eqnarray}
It can be done by expanding the Vandermonde determinant in the sum
over all permutations, evaluating the corresponding integrals and
resuming the resulting expression back to form another Vandermonde
determinant:
\[
I=(-1)^{\frac{n_B(n_B+3)}{2}}\frac{\prod_{k=N-2n_B}^{N-1}k!}
{\left[\prod_{l=1}^{n_B}\left(-iN\mu^{(l)}_{1B}\right)
\left(-iN\mu^{(l)}_{2B}\right)\right]^{N-2n_B+1}}
\,\,\Delta\left\{\left(-iN\mu^{(l)}_{1B}\right)^{-1},
\left(-iN\mu^{(l)}_{2B}\right)^{-1}\right\}
\]
Combining all these facts we restore the normalisation constant
as:

\begin{equation}\label{norcon}
\tilde{C}=\frac{(-1)^{N(n_b+n_F)-n_B(n_B/2-1)}N^{2n_B(N-n_B)+n_B+n_F}}
{(2\pi)^{n_F}\prod_{k=1}^{2n_B}\Gamma(N-k+1)}
\end{equation}
Coming back to investigating the  expression Eq(\ref{kk3}) we can
already  continue analytically: $\mu^{(k)}_{F}\to i\mu^{(k)}_{F}$
for $k=1,...,2n_F$ and set all imaginary parts of the spectral
parameters $\mu^{(l)}_{1B}$  and $\mu^{(l)}_{2B}$ to zero. As
usual, we are interested in the so-called "scaling limit" when all
the spectral parameters $\mu^{(l)}_{1B}$, $\mu^{(l)}_{2B}$ as well
as $\mu^{(k)}_{F}$ are around the same point of the spectrum
 $\mu$ such that $|\mu|<2$,
their mutual distance being of the order of $N^{-1}$.
Correspondingly, we set:
\[
\mu^{l}_{(1,2)B}=\mu+\frac{1}{N}\omega^{(l)}_{B(1,2)}\,,\,
\mu^{(k)}_{F}=\mu+\frac{1}{N}\omega^{(k)}_F
\]
and consider all  $\omega_B\,,\,\omega_F=O(1)$ when $N\to \infty$.

In this way we reduce the expression under investigation to the form
convenient for starting the saddle-point analysis:
\begin{eqnarray}\label{prom2}
\nonumber
{\cal K}_{N}(\hat{\mu}_B,\hat{\mu}_F)&=&\tilde{C}_1
\frac{e^{\frac{N}{2}\mbox{Tr}[\hat{\mu}_F]^2}}
{\Delta\{\hat{\omega}_F\}\Delta\{\hat{\omega}_B\}}
\int_{R}dq_{1} ... \int_{R}dq_{2n_F}
\Delta\{\hat{q}_F\}\left[\prod_{k=1}^{2n_F}q_k\right]^{-2n_B}
e^{i\sum_{k=1}^{2n_F}\omega^{(k)}_F q_{k}}
\\  &\times& e^{-N\left[\sum^{2n_F}_{k=1}{\cal L}_{F}(q_k)\right]}
\\ \nonumber &\times&
\int_{R_{+}}dp^{(1)}_{1} ... \int_{R_{+}}dp^{(n_B)}_1
\int_{R_{+}}dp^{(1)}_2 ... \int_{R_{+}}dp^{(n_B)}_2
\Delta\{\hat{p}_1\}\Delta\{\hat{p}_2\} \prod_{l_1,l_2=1}^{n_B}
\left(p^{(l_1)}_1+p^{(l_2)}_2\right)
\\ \nonumber &\times&
\left[\prod_{l=1}^{n_B}p^{(l)}_1p^{(l)}_2\right]^{-2n_F}
e^{i\sum_{l=1}^{n_F}\left(\omega^{(l)}_{1B} p^{(l)}_1
-\omega^{(l)}_{2B} p^{(l)}_2\right)}
\prod_{k=1}^{2n_F}\prod_{l=1}^{n_B} \left[q_{k}-p^{(l)}_1\right]
\left[q_{k}+p^{(l)}_2\right]
\\ \nonumber &\times&
e^{-N\sum^{n_B}_{l=1}\left[{\cal L}_{1B}(p^{(l)}_1)+
{\cal L}_{2B}(p^{(l)}_2)\right]}
\end{eqnarray}
where
\[
\tilde{C}_1=(-1)^{n_F(n_F-1/2)}N^{n_F(2n_F-1)+n_B(2n_B-1)}\tilde{C}
\]
and
\begin{eqnarray}\label{prom33}
{\cal L}_{F}(q)=\frac{1}{2}q^2-i\mu q-\ln{q}\quad,\quad\\
{\cal L}_{1B}(p)=\frac{1}{2}p^2-i\mu p-\ln{p}\quad,\quad
{\cal L}_{1B}(p)=\frac{1}{2}p^2+i\mu p-\ln{p}
\end{eqnarray}

 Now it is evident that in the limit $N\to \infty$
the contributions to integrals come from the stationary points of
the "actions" ${\cal L}_{F}(q),\,{\cal L}_{1B}(p)$ and ${\cal L}_{1B}(p)$
given by solutions of the equation $q-i\mu-q^{-1}=0$:
\begin{eqnarray}\label{prom44}
q_{k}=\frac{1}{2}\left[i\mu\pm\sqrt{4-\mu^2}\right]\equiv q^{\pm}\quad,
\quad k=1,...,2n_F
\\ \nonumber
p^{(l)}_1=\frac{1}{2}\left[i\mu+\sqrt{4-\mu^2}\right]\equiv q^{+}\quad,\quad
p^{(l)}_2=\frac{1}{2}\left[-i\mu+\sqrt{4-\mu^2}\right]\equiv - q^{-} \,
\quad l=1,...,n_B
\end{eqnarray}
Here we took into account the restrictions of
the original integration domain:
$\mbox{Re} p^{(l)}_{(1,2)}\ge 0$.

Presence of the Vandermonde determinants
as well as the factor $\prod_{k=1}^{2n_F}\prod_{l=1}^{n_B}
\left[q_{k}-p^{(l)}_1\right]
\left[q_{k}+p^{(l)}_2\right]$ makes the integrand
vanish at the saddle-point sets and thus care should
be taken when calculating the saddle point contribution to the
integral. First of all, the totality of $2^{2n_F}$
saddle-points $\hat{q}^{s.p}_F=(q^{\pm}_{1},...,
 q^{\pm}_{2n_F})$ can be further subdivided into
 classes giving contributions of different orders of magnitude
in powers of the small parameter $N^{-1}$.
A little inspection reveals that the leading contribution
comes from the choice of half of saddle-points  to be
$q^{+}$, the rest being $q^{-}$, with total number of such sets
$\left(\begin{array}{c}2n_F\\n_F\end{array}\right)$
(compare \cite{KM,BH}). Indeed, for such a choice the number of
vanishing brackets inside the Vandermonde determinant
$\prod_{1\le k_1<k_2<2n_F}\left(q_{k_1}-q_{k2}\right)$
is minimal.

To find the contribution from each of the relevant saddle-point sets
explicitly let us subdivide the index set $1,2,...,2n_F$ into
the set $\{K_{+}\}=(k_1<k_2<...<k_{n_F}$ of
those indices $1\le k_m\le 2n_F$ for which $q_{k_m}=q^{+}$ and
the rest denoted as $\{K_{-}\}$. Let us also present the integration
variables $q_k$ as : $q_{k\in
\{K_{\pm}\}}=q^{\pm}+\alpha^{\pm}_k$,
with two set of variables $\hat{\alpha}^{\pm}=
\left(\alpha^{\pm}_{k\in\{ K_{\pm}\}}\right) $
 serving to describe deviations from
 the saddle-point values. Then:

\begin{eqnarray}
\nonumber \Delta\{\hat{q}_F\}&=&
\prod_{1\le k_1<k_2<2n_F}\left(q_{k_1}-q_{k_2}\right)
\\
&=&\prod_{\begin{array}{c} k_1\in\{K_{+}\}
\\ k_2\in\{K_{+}\}\end{array}}
\left(q_{k_1}-q_{k_2}\right)
\prod_{\begin{array}{c} k_1\in\{K_{-}\}\\ k_2\in\{K_{-}\}
\end{array}}
\left(q_{k_1}-q_{k_2}\right)
\prod_{\begin{array}{c} k_1\in\{K_{+}\}\\ k_2\in\{K_{-}\}
\end{array}}
\left(q_{k_1}-q_{k_2}\right)
\\ \nonumber &=& (-1)^{\epsilon_{K_{+},K_{-}}}
(4-\mu^2)^{\frac{n^2_F}{2}}\Delta\{\hat{\alpha}^{+}\}
\Delta\{\hat{\alpha}^{-}\}+ \mbox{h. o. t.}  \quad,
\end{eqnarray}
where ${\epsilon_{K_{+},K_{-}}}$ is odd or even integer
serving to take into account the sign factor arising in the process
of rearranging indices in the last of three products in the above
equation. The abbreviation h.o.t. stands for
higher order terms in $\alpha$'s.
Further we expand in the exponentials up to terms
quadratic with respect to $\alpha$'s and
have:

\begin{eqnarray}
e^{i\sum_{k=1}^{2n_F}\omega^{(k)}_F q_{k}}
&=&
e^{iq^{+}\sum_{k\in \{K_{+}\}}\omega^{(k)}_F
 +iq^{-}\sum_{k\in \{K_{-}\}}\omega^{(k)}_F }
e^{i\sum_{k\in \{K_{+}\}}\omega^{(k)}_F \alpha^{+}_{k}
+i\sum_{k\in \{K_{-}\}}\omega^{(k)}_F \alpha^{-}_{k}}
\\
 e^{-N\left[\sum^{2n_F}_{k=1}{\cal L}_{F}(q_k)\right]}&=&
(-1)^{Nn_F}e^{-Nn_F(1+\frac{\mu^2}{2})-
\frac{N}{2}\sqrt{4-\mu^2}\left(q^{+}\sum_{k\in \{K_{-}\}}
\left(\alpha^{-}_k\right)^2-q^{-}\sum_{k\in \{K_{+}\}}
\left[\alpha^{+}_k\right]^2\right)}
\end{eqnarray}
where we made use of $q^{+}q^{-}=-1\,,\, q^{+}+q_{-}=i\mu\,,\,
\frac{1}{2}\left([q^{+}]^2+[q^{-}]^2\right)=1-\mu^2/2$
and $1+\frac{1}{[q^{\pm}]^2}
=\mp q^{\mp}\sqrt{4-\mu^2}$.

Similarly, we set $p^{(l)}_1= q^{+}+\beta^{+}_{l}\quad,\quad
p^{(l)}_2= -(q^{-}+\beta^{-}_{l})$ to describe deviations of
$p^{(l)}_{(1,2)}$ from their saddle-point values. In this way we obtain:
\begin{eqnarray}
&&\prod_{k=1}^{2n_F}\prod_{l=1}^{n_B}
\left[q_{k}-p^{(l)}_1\right]
\left[q_{k}+p^{(l)}_2\right]\\
\nonumber &=& \left[-\left(4-\mu^2\right)\right]^{n_Bn_F}
\prod_{l=1}^{n_B}\prod_{k\in \{K_{+}\}}
\left[\alpha^{+}_{k}-\beta^{+}_{l}\right]
\prod_{k\in \{K_{+}\}}\left[\alpha^{-}_{k}-\beta^{-}_{l}\right]+\mbox{h. o. t.}
\\
&&\Delta\{\hat{p}_1\}\Delta\{\hat{p}_2\} \prod_{l_1,l_2=1}^{n_B}
\left(p_{1,l_1}+p_{2,l_2}\right) =(-1)^{\frac{n_B}{2}}
\left[-\left(4-\mu^2\right)\right]^{\frac{n^2_B}{2}}\Delta\{\hat{\beta}^{+}\}
\Delta\{\hat{\beta}^{-}\}+\mbox{h. o. t.}
\end{eqnarray}
and
\begin{eqnarray}
e^{i\sum_{l=1}^{n_B}\left(\omega^{(l)}_{1B} p^{(l)}_1
-\omega^{(l)}_{2B} p^{(l)}_2\right)}&=&
e^{iq^{+}\sum_{l=1}^{n_B}\omega^{(l)}_{1B}
+iq^{-}\sum_{l=1}^{n_B}\omega^{(l)}_{2B}}
e^{i\sum_{l=1}^{n_B}\left(\omega^{(l)}_{1B} \beta^{+}_{l}
+\omega^{(l)}_{2B} \beta^{-}_{l}\right)}
\\ \nonumber
e^{-N\sum^{n_B}_{l=1}\left[{\cal L}_{B}(p^{(l)}_1)+{\cal L}_{B}(p^{(l)}_2)
\right]}&=&
e^{-Nn_B(1+\frac{\mu^2}{2})-\frac{N}{2}\sqrt{4-\mu^2}\left(\sum_{l}q^{+}
\left[\beta^{-}_{l}\right]^2-\sum_{l}q^{-}
\left[\beta^{+}_{l}\right]^2\right)}
\end{eqnarray}

Let us now introduce four diagonal matrices of size $n=n_B+n_F$:
\begin{eqnarray}
\hat{\Theta}^+&=&\mbox{diag}\left(\hat{\alpha}^{+},\hat{\beta}^{+}\right)\quad,\quad
\hat{\Theta}^-=\mbox{diag}\left(\hat{\alpha}^{-},\hat{\beta}^{-}\right)
\\ \nonumber &&
\hat{\Omega}^-=\mbox{diag}\left(\hat{\omega}^{-}_F,\hat{\omega}^{-}_B\right)
\quad,\quad \hat{\Omega}^+=
\mbox{diag}\left(\hat{\omega}^{+}_F,\hat{\omega}^{+}_B\right)
\end{eqnarray}
where $\hat{\omega}^{\pm}_F=\mbox{diag}\left(\omega^{(k)}_F: \, k\in
\{K_{\pm}\}\right)$ and $\hat{\omega}^{+}_B=
\mbox{diag}\left(\omega^{(l)}_{1B}\right)
\,,\, \hat{\omega}^{-}_B=
\mbox{diag}\left(\omega^{(l)}_{2B}\right)\,,\, l=1,...,n_B$.

Collecting all the factors we now can represent the
leading order contribution to
the correlation function as $N\gg 1$ in the form:
\begin{eqnarray}\label{qq1}
\nonumber
&& {\cal K}_{N}(\hat{\mu}_B,\hat{\mu}_F)=
{\cal K}_{s.p.}(\hat{\mu}_B,\hat{\mu}_F)\\
&\times &\int_{R^{n}}d\hat{\Theta^{+}}\Delta\left\{\hat{\Theta}^{+}\right\}
\exp\left\{-\frac{N}{2}q^+\sqrt{4-\mu^2}\mbox{Tr}\left[\hat{\Theta}^{+}\right]^2
+i\mbox{Tr}\left[\hat{\Theta}^{+}\hat{\Omega}^+\right]\right\}
\\ \nonumber &\times& \int_{R^{n}}d\hat{\Theta}^{-}
\Delta\left\{\hat{\Theta}^{-}\right\}
\exp\left\{\frac{N}{2}q^{-}\sqrt{4-\mu^2}
\mbox{Tr}\left[\hat{\Theta}^{-}\right]^2
+i\mbox{Tr}\left[\hat{\Theta}^{-}\hat{\Omega}^-\right]\right\}
\\ &=& \frac{(2\pi)^n}{N^{n^2}}(-1)^{-n/2} {\cal K}_{s.p.}(\hat{\mu}_B,\hat{\mu}_F)
(4-\mu^2)^{-\frac{n^2}{2}}
\Delta\left\{\hat{\Omega}^{-}\right\}\Delta\left\{\hat{\Omega}^{+}\right\}
\\ \nonumber &\times&
\exp\left\{\frac{1}{2N\sqrt{4-\mu^2}}\left(\frac{1}{q^-}
\mbox{Tr}\left[\hat{\Omega}^{-}\right]^2
-\frac{1}{q^+}\mbox{Tr}\left[\hat{\Omega}^+\right]^2\right)\right\}
\end{eqnarray}
where
\begin{eqnarray}\label{qq2}
{\cal K}_{s.p.}(\hat{\mu}_B,\hat{\mu}_F)&\propto&
\frac{e^{Nn_F\mu^2 +\mu \mbox{Tr}\hat{\omega}_F+
\frac{1}{2N}\mbox{Tr}\hat{\omega}^2_F}}
{\Delta\{\hat{\omega}_F\}\Delta\{\hat{\omega}_B\}}
(-1)^{\tilde{\epsilon}}\tilde{C}_1 (4-\mu^2)^{\frac{n^2}{2}}e^{-Nn
\left(1+\frac{\mu^2}{2}\right)}
\\ \nonumber &\times&
e^{iq^{+}\left[\sum_{k\in\{K_{+}\}}\omega^{(k)}_{F}+
\sum_{l=1}^{n_B}\omega^{(l)}_{1B}\right]+
iq^{-}\left[\sum_{k\in\{K_{-}\}}\omega^{(k)}_{F}+
\sum_{l=1}^{n_B}\omega^{(l)}_{2B}\right]}
\end{eqnarray}
Here we used the integral formula
\begin{eqnarray}
\int_{R^{m}}d\hat{\Theta}\Delta\left\{\hat{\Theta}\right\}
\exp\left\{-\frac{t}{2}\mbox{Tr}\left[\hat{\Theta}\right]^2
+i\mbox{Tr}\left[\hat{\Theta}\hat{\Omega}\right]\right\}
=\frac{(-1)^{\frac{m(m-1)}{4}}(2\pi)^{m/2}}
{t^{\frac{m^2}{2}}}\Delta\left\{\hat{\Omega}\right\}
\exp\left\{-\frac{1}{2t}\mbox{Tr}\left[\hat{\Omega}\right]^2\right\}
\end{eqnarray}
and denoted: $\tilde{\epsilon}=\epsilon_{K_{+},K_{-}}+Nn_F+n_Bn_F
+n_B/2$.

A close inspection of the quotient of the Vandermonde determinants
occurring when substituting Eq.(\ref{qq2}) into Eq.(\ref{qq1})
reveals that:
\begin{eqnarray}
\frac{\Delta\left\{\hat{\Omega}^{-}\right\}
\Delta\left\{\hat{\Omega}^{+}\right\}}
{\Delta\{\hat{\omega}_F\}\Delta\{\hat{\omega}_B\}}
=(-1)^{\epsilon_{K_{+},K_{-}}}F^{K_{+},K_{-}}_{n_B,n_F}
\left(\hat{\omega}_B,\hat{\omega}_F\right)
\end{eqnarray}
where we introduced the notation:
\begin{eqnarray}
F^{K_{+},K_{-}}_{n_B,n_F}
\left(\hat{\omega}_B,\hat{\omega}_F\right)=
\displaystyle{\frac{\prod_{l=1}^{n_B}\left[
\prod_{k\in \{K_{+}\}}\left(\omega_F^{(k)}-\omega^{(l)}_{1B}\right)
\prod_{k\in \{K_{-}\}}\left(\omega_F^{(k)}-\omega^{(l)}_{2B}\right)\right]}
{\prod_{l_1<l_2}^{n_B}
\left(\omega_{1B}^{(l_1)}-\omega^{(l_2)}_{2B}\right)
\prod_{\begin{array}{c}k_1\in \{K_{-}\}\\k_2\in \{K_{+}\}\end{array}}
\left(\omega_F^{(k_1)}-\omega^{(k_2)}_{F}\right)}}
\end{eqnarray}
and $(-1)^{\epsilon_{K_{+},K_{-}}}$ is exactly the same factor that
appeared in our calculation earlier due to rearranging
variables inside the brackets in the product of Vandermonde determinants.

We also observe that:
\[
\mu\mbox{Tr}\hat{\omega}_F+iq^{+}\sum_{k\in \{K_{+}\}}\omega^{(k)}_{F}
+iq^{-}\sum_{k\in \{K_{-}\}}\omega^{(k)}_{F}
=-iq^{+}\sum_{k\in \{K_{-}\}}\omega^{(k)}_{F}
-iq^{-}\sum_{k\in \{K_{+}\}}\omega^{(k)}_{F}
\]
The last relation allows us to write down the final result
of the calculation in the form:
\begin{eqnarray}
{\cal K}_{N\to \infty}(\hat{\mu}_B,\hat{\mu}_F)
&&=C_{N,n_B,n_F}e^{\frac{N}{2}
(n_F-n_B)\mu^2}\\ \nonumber &\times&
\sum_{\{K_{+}\}}
F^{K_{+},K_{-}}_{n_B,n_F}\left(\hat{\omega}_B,\hat{\omega}_F\right)
e^{iq^+\left[\sum_{l=1}^{n_B}\omega^{(l)}_{1B}-\sum_{k\in
\{K_{-}\}}\omega^{(k)}_{F}\right]+
iq^-\left[\sum_{l=1}^{n_B}\omega^{(l)}_{2B}-\sum_{k\in
\{K_{+}\}}\omega^{(k)}_{F}\right]}
\end{eqnarray}
and the summation goes over all possibilities of subdividing
the index set $1,2,...,2n_F$ into two index sets $\{K_{+}\}$ and
$\{K_{-}\}$. Here $C_{N,n_B,n_F}$ stands for the overall
normalisation constant:
\[
C_{N,n_B,n_F}=(2\pi)^{n_B}(-1)^{Nn_B+n_F^2-n_F+\frac{n_B^2}{2}
+n_B+n_Bn_F}
\]
 and we neglected
 all the terms of the order of $O(N^{-1})$ in the exponential
to be consistent with the leading order approximation.

Remembering that the mean spectral density of the GUE eigenvalues
in the limit of large $N$ is given by the Wigner semicircular law:
$\rho(\mu)=\frac{1}{2\pi}\sqrt{4-\mu^2}$
so that $q^{\pm}=\frac{i\mu}{2}\pm \pi\rho(\mu)$ we satisfy ourselves that
for $\mu=0$ the derived expression coincides with one announced
in \cite{AS} and obtained by a rather different method.
\section{Conclusions}
In the present paper we have demonstrated that the method
suggested in [I] allows one to analyse the correlation function
containing both positive and negative moments of characteristic
polynomials. This technique combines simultaneous exploitation of
the standard Hubbard-Stratonovich transformation with an
integration theorem  (see formula (\ref{trans})). The latter is a
new element as compared with [I] which replaces the Ingham-Siegel
integration used there. The method leads to a compact integral
representation (\ref{prom3}) for the correlation function. To
study the asymptotic limit of large GUE matrices we needed to
expose variables amenable to the saddle point treatment. For this
purpose  we had to derive the integration formula extending the
Itzykson-Zuber-Harish-Chandra integral to the non-compact K\"ahler
manifold $\sf{U(n_1,n_2)/\sf{T}}$ (see expression
(\ref{IZpseudo})). Our derivation is based on the
Duistermaat-Heckman localization principle.

Our preliminary  considerations show that the outlined procedure
works well for other ensembles of random matrices, in
particular for the chiral GUE and non-Hermitian ensembles.
In the latter cases it
requires a non-compact analogue of the integral formula found by
Guhr, Wettig \cite{guhr} and by Jackson, Sener and Verbaarschot
\cite{verbaarschot}. The corresponding calculation will be
published elsewhere.

\section{Acknowledgements}
E Strahov is grateful to R Picken for useful communications.
Y V Fyodorov would like to thank B Duplantier for pointing out
the reference \cite{Dupl} in relation to
the formula Eq.(\ref{trans1}), to DW Farmer for informing him on the paper
\cite{Farmer}, to G Akemann for bringing references
\cite{DN,Szabo1} and to P.-E. Paradan for bringing references
\cite{rossmann,berline,paradan} to the authors attention.

This
research was supported by EPSRC grant GR/13838/01 "Random matrices
close to unitary or hermitian."

\section*{Appendix A. Diffusion derivation of the Itzykson-Zuber
type integral on the pseudo-unitary group $\sf{U(n_1,n_2)}$} Let
us consider a diffusion on matrices that are elements of the Lie
algebra $\sf{u(n_1,n_2)}$. Any such $\sf{(n_1+n_2)\times
(n_1+n_2)}$ matrix $A$ satisfies the equation:
\begin{equation}\label{AIZ1}
A^{\dagger}=\lambda A \lambda
\end{equation}
as it follows from the pseudo-unitarity of the group $\sf
{U(n_1,n_2)}$. The corresponding Laplace operator invariant under
pseudo-unitary transformations acquires the following form:
\begin{equation}\label{AIZ2}
D_{A}=\sum\limits_{i=1}^{n_1+n_2}\partial^2/\partial
A_{ii}^2+1/2\sum\limits_{1\leq i<j\leq n_1+n_2} (-)^{\sigma_{ij}}
\left[\partial^2/(\partial
\mbox{Re}\;A_{ij})^2+\partial^2/(\partial
\mbox{Im}\;A_{ij})^2\right]
\end{equation}
where the symbol $\sigma_{ij}$ takes the values $0$  when $1\leq
i<j\leq {\sf{n_1}} $ or ${\sf{n_1}}<i<j\leq {\sf{n_2}}$ and $1$
when $1\leq i\leq {\sf{n_1}}<j\leq {\sf{n_2}}$.
 Once the Laplacian $D_A$ is given, it generates the diffusion on
 matrices satisfying Eq.(\ref{AIZ1}). Such a diffusion is described by
 the heat equation
\begin{equation}\label{AIZ3}
\frac{1}{2}\;D_A\psi(A,t)=\partial_t\psi(A,t)
\end{equation}
with the initial condition
\begin{equation}\label{AIZ4}
\psi(A,t=0)=\phi(A)
\end{equation}
The solution of the diffusion problem defined above is
represented by an integral over the matrices $B$ satisfying the
condition Eq.(\ref{AIZ1}):
\begin{equation}\label{AIZ5}
\psi(A,t)=\frac{1}{(2\pi t)^{({\sf{n_1+n_2}})^2/2}}\int
dB\exp\left(-\frac{1}{2t}\mbox{Tr}(A-B)^2\right)\phi(B)
\end{equation}
Let us note that the expression $\mbox{Tr}(A-B)^2$ is not positive definite for
the matrices from the Lie algebra $\sf{u(n_1,n_2)}$. However one can
always choose the initial distribution $\phi(B)$ in
 a way that ensures the existence of the integral Eq.(\ref{AIZ5}). Let us further assume
 that the initial distribution $\phi(B)$ is invariant under
pseudo-unitary transformations:
\begin{equation}\label{AIZ6}
\phi(gBg^{-1})=\phi(B),\;\; g\in\sf{U(n_1,n_2)}
\end{equation}
This condition implies that the function $\phi(B)$ depends only
on the set eigenvalues of the matrix $B$ and is symmetric under separate permutations of
the first $\sf{n_1}$ and the rest of $\sf{n_2}$ eigenvalues. Indeed,
the corresponding Weyl group is $\sf{S_{n_1}\times S_{n_2}}$ and
not $\sf{S_{n_1+n_2}}$ as in the case of unitary transformations.

In what follows we adopt the argumentation used in the original work by
Itzykson and Zuber \cite{IZ} to the present case. Let us take a diagonal matrix $A$
in Eq. (\ref{AIZ5}) and diagonalize the matrix $B$ by a
pseudo-unitary transformation:
\begin{equation}\label{AIZ7}
A=\mbox{diag}\left(\alpha_1,\alpha_2,\cdots
,\alpha_{{\sf{n_1+n_2}}}\right),\;\;\;
B=v\;\mbox{diag}\left(\beta_1,\beta_2,\cdots
,\beta_{{\sf{n_1+n_2}}}\right)v^{-1},\;\;\;v\in{\sf{U(n_1,n_2)}}
\end{equation}
Then the integral expressing the solution of the diffusion
problem (Eq.(\ref{AIZ5})) can be rewritten as follows:
\begin{equation}\label{AIZ8}
\psi(\alpha,t)=\frac{1}{(2\pi t)^{{\sf{(n_1+n_2)}}^2/2}}\int
dv\int d\beta
\triangle^2(\beta)\exp\left(-1/2t\;\mbox{Tr}\left[(\alpha-v\beta
v^{-1})^2\right]\right)\phi(\beta)
\end{equation}
where $\alpha,\beta$ stand for the diagonal matrices
$\mbox{diag}\left(\alpha_1,\alpha_2,\cdots
,\alpha_{{\sf{n_1+n_2}}}\right)$ and  $\mbox{diag}\left(\beta_1,\beta_2,\cdots
,\beta_{{\sf{n_1+n_2}}}\right)$, respectively. We introduce the function
$\zeta(\alpha ,t)$ antisymmetric with respect to separate permutations
inside the sets $\left(\alpha_1,\alpha_2,\cdots
,\alpha_{{\sf{n_1}}}\right)$ and
$\left(\alpha_{{\sf{n_1}}+1},\alpha_{{\sf{n_1}}+2},\cdots
,\alpha_{{\sf{n_1+n_2}}}\right)$:
\begin{equation}\label{AIZ9}
\zeta(\alpha,
t)=\triangle(\alpha)\psi(\alpha,t),\;\;\;\xi(\alpha,t=0)=\triangle(\alpha)\phi(\alpha)
\end{equation}
Act on this function by the Laplacian operator $D_A$.
For the function $\zeta(\alpha,t)$ depending only on the eigenvalues
of the matrix $A$ and $\psi(\alpha,t)$ being a solution of the
diffusion problem Eq.(\ref{AIZ3}), the procedure yields the following differential
equation:
\begin{equation}\label{AIZ10}
\partial_t\xi(\alpha,t)=1/2\;
\sum\limits_{i=1}^{n_1+n_2}\frac{\partial^2}{\partial\alpha^2_i}\xi(\alpha,t)
\end{equation}
The only diffusion kernel ${\sf{K}}(\alpha ,\beta ,t)$ corresponding to the above equation
which is antisymmetric with respect to separate permutations inside the sets
$\left(\beta_1,\beta_2,\cdots ,\beta_{{\sf{n_1}}}\right)$ and
 $\left(\beta_{{\sf{n_1}}+1},\beta_{{\sf{n_1}}+2},\cdots
,\beta_{{\sf{n_1+n_2}}}\right)$ is given by
\begin{equation}\label{AIZ11}
{\sf{K}}(\alpha ,\beta ,t)=\frac{\mbox{const}}{(2\pi
t)^{({\sf{n_1+n_2}})/2}}\sum\limits_{P\in {\sf{S_{n_1}\times
S_{n_2}}}}(-)^P\exp\left[-\frac{1}{2t}\sum\limits_{i}(\alpha_i-\beta^P_i)^2\right]
\end{equation}
Comparison of the equations Eqs.(\ref{AIZ8},\ref{AIZ9}) and Eq.(\ref{AIZ11}) yields the desired formula
Eq.(\ref{IZpseudo}) after a simple manipulation.
\section*{Appendix B. Complex parameterization of
${\sf{U(2)/U(1)\times U(2)}}$
and ${\sf{U(1,1)/U(1)\times U(1)}}$}

First we note that the manifold
$\sf{U(2)/U(1)\times U(1)}$ is equivalent to $\sf{SU(2)/U(1)}$.
The complex Lie algebra
corresponding to the group $\sf{SU(2)}$
has three elements $e_q, e_{-q}, h$ in its Cartan-Weil basis.
In the two-dimensional fundamental representation
these basis elements are expressed as follows:
\begin{equation}\label{a11}
h=1/2\left(\begin{array}{cc}
  1 & 0 \\
  0 & -1
\end{array}\right);\;\;\;
e_q=\left(\begin{array}{cc}
  0 & 1 \\
  0 & 0
\end{array}\right);\;\;\;
e_{-q}=\left(\begin{array}{cc}
  0 & 0 \\
  -1 & 0
\end{array}\right)
\end{equation}
Following the general method of constructing the complex
parameterization we decompose the
representative $g(z,\bar z)$ of the coset
space $\sf{SU(2)/U(1)}$ in the same way as it is done in the formula
Eq.(\ref{gcdecomposition}). The factors $u(z)$ and $p(z,\bar z)$
are given by:
\begin{equation}\label{a12}
u(z)=\exp\left(ze_q\right),\;\;\;p(z,\bar z)=\exp\left(y(z,\bar
z)e_q\right)\cdot\exp\left(k(z,\bar z)h\right)
\end{equation}
Any element of a coset of a unitary group must satisfy the
unitarity condition. For our case the unitarity condition
$g^{\dag}(z,\bar z)=g^{-1}(z,\bar z), \forall g(z,\bar
z)\in\sf{SU(2)/U(1)}$ is equivalent to the following algebraic
relation:
\begin{equation}\label{a13}
p(z,\bar z)p^{\dagger}(z,\bar
z)=u^{-1}(z)\left(u^{\dagger}(z)\right)^{-1}
\end{equation}
This relation enables one to derive explicitly
the functions $y(z,\bar z)$ and $k(z,\bar z)$ entering the formula
Eq.(\ref{a12}). We chose to perform the calculations below
 in the fundamental matrix representation since the obtained expressions for
$y(z,\bar z)$ and $k(z,\bar z)$ are the same in any
representation. In the two-dimensional matrix
representation we have:
\begin{eqnarray}\label{a14}
u(z)=\left(\begin{array}{cc}
  1 & z \\
  0 & 1
\end{array}\right),\qquad\qquad\;
\exp\left(y(z,\bar z)e_{-q}\right)= \left(\begin{array}{cc}
  1 & 0 \\
  -y(z,\bar z) & 1
\end{array}\right)\nonumber \\
\exp\left(k(z,\bar z)h\right)= \left(\begin{array}{cc}
  \exp\left(1/2k(z,\bar z)\right) & 0 \\
  0 & \exp\left(-1/2k(z,\bar z)\right)
\end{array}\right)\qquad
\end{eqnarray}
Inserting the above matrix expressions to the formula (\ref{a13})
we find:
\begin{equation}
Re\left(k(z,\bar z)\right)=\ln(1+z\bar z),\;\;\;y(z,\bar z)=\bar
z/(1+z\bar z)
\end{equation}
Let us note that the function $k(z,\bar z)$ is specified up to an
arbitrary complex part. This means that the corresponding
 element of the coset space $g(z,\bar z)$ is determined up
to a multiplication by a torus element from the right,
as it must be the case.

The parameterization of the coset space $\sf{U(1,1)/U(1)\times
U(1)}$ is obtained by following similar steps. The difference is that the
corresponding representative of the coset space should be an element
of the pseudo-unitary group $\sf{U(1,1)}$ rather than of $\sf{U(2)}$.
Therefore the representative $g(z,\bar z)$ of the coset
space $\sf{U(1,1)/U(1)\times U(1)}$ must  satisfy the pseudo-unitary
condition Eq.(\ref{pseudounitarygroup}). The latter leads to the
following algebraic relation (which replaces Eq.(\ref{a13}) above):
\begin{equation}
p(z,\bar z)\lambda p^{\dagger}(z,\bar z)= u^{-1}(z)\lambda
(u^{\dagger})^{-1}
\end{equation}
Using this formula we obtain the expressions for the real
part of the function $k(z,\bar z)$ and for the function $y(z,\bar
z)$:
\begin{equation}
Re\left(k(z,\bar z)\right)=\ln(1-z\bar z),\;\;\;y(z,\bar z)=-\bar
z/(1-z\bar z)
\end{equation}
A remarkable feature of the described parameterization is that the
expression for $Re\left(k(z,\bar z)\right)$ can be considered as
fundamental K\"ahler potentials. This is a quite general property
common to any homogeneous K\"ahler manifold with a unitary or
pseudo-unitary transformation group as was shown by Itoh, Kugo and
Kunitomo \cite{itoh}

An alternative way to find the K\"ahler potentials in terms of
the local
complex parameters $z,\bar z$ is to exploit the relation
Eq.(\ref{funitarykpotential}) (or its analogue
Eq.(\ref{kahlerpseudo}) for a
pseudo-unitary coset). The number of fundamental K\"ahler
potentials is equal to the number of projection matrices or to the rank
of the Lie algebra under consideration. The Lie algebra $su(2)$ has
one basis element $h$ in its Cartan subalgebra, so the rank of
$su(2)$ is equal to unity. The projection matrix corresponding to
the basis element $h$ is determined from the equations
(\ref{projection}):
\begin{eqnarray}
\eta=\left(\begin{array}{cc}
  0 & 0 \\
  0 & 1
\end{array}\right)
\end{eqnarray}
Now insert the projection matrix $\eta$ and the matrix $u(z)$
given by Eq.(\ref{a14}) to the formula
Eq.(\ref{funitarykpotential}) for the unitary coset space
$\sf{U(2)/U(1)\times U(1)}$ ( or to its counterpart
Eq.(\ref{kahlerpseudo} for the pseudo-unitary coset
$\sf{U(1,1)/U(1)\times U(1)}$). A simple calculation yields the
corresponding K\"ahler potentials:
\begin{equation}
K_{{\sf U(2)/U(1)\times U(1)}}(z,\bar z)=\ln(1+z\bar z),\;\;\;
K_{{\sf U(1,1)/U(1)\times U(1)}}(z,\bar z)=\ln(1-z\bar z)
\end{equation}
Once the K\"ahler potentials are known, the (1.1) forms on the
manifolds $\sf{U(2)/U(1)\times U(1)}$ and $\sf{U(1,1)/U(1)\times
U(1)}$ can be immediately obtained from the relation
Eq.(\ref{oneoneform}):
\begin{equation}
\Omega_{{\sf U(2)/U(1)\times U(1)}}=-\frac{1}{2\pi
i}\frac{dz\wedge d\bar z}{(1+z\bar z)^2},\;\;\; \Omega_{{\sf
U(1,1)/U(1)\times U(1)}}=\frac{1}{2\pi i}\frac{dz\wedge d\bar
z}{(1-z\bar z)^2}
\end{equation}
Finally let us determine the momentum maps on the manifolds
$\sf{U(2)/U(1)\times U(1)}$ and $\sf{U(1,1)/U(1)\times U(1)}$. As
it can be seen from Eq.(\ref{rhotrace}) the momentum maps
are completely determined by the matrix $\rho(z,\bar z)$. In order to
find that matrix we insert the projection matrix $\eta$ and the
matrix $u(z)$ to the formula Eq.(\ref{rhoexpression}) for the compact
coset $\sf{U(2)/U(1)\times U(1)}$ and to the formula
Eq.(\ref{projpseudo}) for its non-compact counterpart
$\sf{U(1,1)/U(1)\times U(1)}$. We obtain:
\begin{equation}
\rho_{{\sf U(2)/U(1)\times U(1)}}(z,\bar z)=\frac{1}{1+z\bar z}
\left(\begin{array}{cc}
  z\bar z & z \\
  z & 1
\end{array}\right),\;\;\rho_{{\sf U(1,1)/U(1)\times U(1)}}(z,\bar z)=\frac{-1}{1-z\bar z}
\left(\begin{array}{cc}
  z\bar z & z \\
  z & -1
\end{array}\right)
\end{equation}
Now the momentum maps $T_q(z,\bar z), T_{-q}(z,\bar z), T_{h}(z,\bar
z) $ corresponding to the basis elements $e_q, e_q$ and $h$ can
be easily constructed. For the space $\sf{U(2)/U(1)\times
U(1)}$ they are given by
\begin{equation}
T_q(z,\bar z)=-\frac{\bar z}{1+z\bar z},\;\; T_{-q}(z,\bar
z)=\frac{z}{1+z\bar z},\;\; T_h(z,\bar z)=\frac{1-z\bar z}{1+z\bar
z}
\end{equation}
The corresponding momentum maps for the non-compact coset
$\sf{U(1,1)/U(1)\times U(1)}$ are
\begin{equation}
T_q(z,\bar z)=\frac{\bar z}{1+z\bar z},\;\; T_{-q}(z,\bar
z)=-\frac{z}{1+z\bar z},\;\; T_h(z,\bar z)=\frac{1+z\bar
z}{1-z\bar z}
\end{equation}
\section*{Appendix C. Transformation of projection matrices $\rho_i(z,\bar z)$}
In order to prove Eq.(\ref{rhotransformation}) we note that
the decomposition $gu(z)=u(gz)p(z,g)$ that defines the group
action on the flag manifold under consideration leads to the
following expressions for $u(gz)$ and $u^{\dag}(gz)$:
\begin{equation}\label{tdecompostransformed}
u(gz)=gu(z)p^{-1}(z,g),\;\;\;
u^{\dag}(gz)=\left(p^{\dag}(z,g)\right)^{-1}u^{\dag}(z)g^{\dag}
\end{equation}
Rewrite $\rho_i(gz,\overline{gz})$ explicitly using formula
Eq.(\ref{rhoexpression}):
\begin{equation}\label{rhotransformedexpression}
\rho_i(gz,\overline{gz})=u(gz)\eta_i\left(\eta_iu^{\dag}(gz)u(gz)\eta_i+I-\eta_i\right)^{-1}\eta_iu^{\dag}(gz)
\end{equation}
Insert $u(gz)$ and $u^{\dag}(gz)$ given by
Eq.(\ref{tdecompostransformed}) to the above formula  and use the
properties of the projection matrices $\eta_j$
(equation (\ref{projection})). Taking into account relations
\begin{equation}\label{projectionrelations}
\eta_i\left(I-\eta_i\right)=0,\;\;\;
\left(I-\eta_i\right)\left(I-\eta_i\right)=\left(I-\eta_i\right)
\end{equation}
we obtain:
\begin{eqnarray}\label{rhoi}
\rho_i(gz,\overline{gz})=\nonumber\qquad\qquad\qquad\qquad\qquad \\
gu(z)p^{-1}(z,g)\eta_i
\left(\eta_ip^{-1}(z,g)\eta_i+I-\eta_i\right)^{-1}\times
\nonumber\qquad\qquad
\\
 \left(\eta_iu^{\dag}(z)u(z)\eta_i+I-\eta_i\right)^{-1}
\left(\eta_i\left(p^{\dag}(z,g)\right)^{-1}\eta_i+I-\eta_i\right)^{-1}
\eta_i\left(p^{\dag}(z,g)\right)^{-1}u^{\dag}(z)g^{\dag}
\end{eqnarray}
The transformation law for the matrices $\rho_i(z,\bar z)$
(see Eq.(\ref{rhotransformation})) follows immediately when we
simplify the above expression using
\begin{eqnarray}
p^{-1}(z,g)\eta_i\left(\eta_ip^{-1}(z,g)\eta_i+I-\eta_i\right)^{-1}=\eta_i,\quad
\nonumber\\
\left(\eta_i\left(p^{\dag}(z,g)\right)^{-1}\eta_i+I-\eta_i\right)^{-1}
\eta_i\left(p^{\dag}(z,g)\right)^{-1}=\eta_i
\end{eqnarray}

\section*{Appendix D. Proof of the Theorem I}

In this Appendix we give a proof of the statement of the
Theorem I. In fact, we demonstrate the validity of the closely related

{\bf Theorem Ia}

{\sf
Consider a function $ F({\bf S_1},...,{\bf S_m})$
of $N$-component real vectors ${\bf S}_l\,\, 1\le l\le m$ such that
\begin{equation}\label{conv1}
\int_{R^N}d{\bf S}_1...\int_{R^N}d^2{\bf S}_m
|F({\bf S_1},...,{\bf S_m})|<\infty
\end{equation}
Denoting ${\bf S}^T$ the transposition
suppose further that the function $F$
depends only on $m(m+1)/2$ scalar products ${\bf S}^{T}_{l_1}
{\bf S}_{l_2}\,\, 1\le l_1,l_2\le m$ so that it can be rewritten as
a function ${\cal F}(\tilde{Q}_m)$ of
$m\times m$ real symmetric matrix $\tilde{Q}_m$:
\[
(\tilde{Q}_m)_{k,l}
=\left({\bf S}^{T}_{k}{\bf S}_{l}\right)\left.\right|_{1\le k,l\le m}
\]
Then for $N>m$ the integral defined as
\begin{equation}\label{defint}
I^{(F)}_{N,M}=\int_{R^N}d{\bf S}_1... \int_{R^N}d{\bf S}_m
F({\bf S}_1,...,{\bf S}_m)
\end{equation}
is equal to
\begin{equation}\label{trans1}
I^{({\cal F})}_{N,M}={\cal C}^{(o)}_{N,m}
\int_{\hat{Q}_m>0} d\hat{Q}_m\left(\mbox{det}\hat{Q}_m\right)^{(N-m-1)/2}
{\cal F}(\hat{Q}_m)
\end{equation}
where the proportionality constant is given by
\[
{\cal C}^{(o)}_{N,m}= \frac{\pi^{\frac{m}{2}\left(N-\frac{m-1}{2}\right)}}
{\prod_{k=0}^{m-1}\Gamma\left(\frac{N-k}{2}\right)}
\]
and the integration in  Eq.(\ref{trans1}) goes over the manifold
of real symmetric positive definite $m\times m$ matrices
$\hat{Q}_m$. }

{\bf Proof}

We prove the statement by induction in $m$ for any $N>m$.

First, for $m=1$ we parameterize ${\bf S}=r\left({\cal O}_N{\bf
e}\right)$ , where $r=||{\bf S}||\ge 0$, the matrix ${\cal O}_N\in
O(N)$ is a real orthogonal $N\times N$ satisfying: ${\cal
O}^T_N{\cal O}_N={\bf 1}_N$ and $N-1$ first components of the
vector ${\bf e}$ are chosen to be zero, the last component being
unity: ${\bf e}^T=(0,...,0,1)$. The integration measure can be
written as $d{\bf S}=r^{N-1}dr d\mu\left({\cal O}\right)$, the
last factor standing for the Haar's measure on the group $O(N)$,
such that:
\[
\int_{O(N)}d\mu\left({\cal O}\right)=
2\frac{\pi^{N/2}}{\Gamma(N/2)}\equiv \Omega_N
\]
Now for $N\ge 2$ we have:
\[
\int_{R^N}d{\bf S}F\left({\bf S}^2\right)=\Omega_N\int_{R^+}
dr r^{N-1}F(r^2)=\frac{\Omega_N}{2}\int_{R^+} dq_{11}
q^{(N-2)/2}_{11}{\cal F}(q_{11})
\]
which proves the statement and gives the value
${\cal C}_{N,1}^{(o)}=\frac{1}{2}\Omega_N$ as required.

Suppose now that the statement is true for $(m-1)$ vectors,
 each with $(N-1)$ real components, that means:
\begin{equation}\label{ind}
I^{(F)}_{N-1,m-1}=I^{({\cal F})}_{N-1,m-1}={\cal C}^{(o)}_{N-1,m-1}
\int_{\hat{Q}_{m-1}>0} d\hat{Q}_{m-1}\left(\mbox{det}
\hat{Q}_{m-1}\right)^{(N-m-1)/2}
{\cal F}\left[\hat{Q}_{m-1}\right]
\end{equation}

To consider the case of $m$ vectors, each with $N$ components we
represent the $m\times m$ matrix $\tilde{Q}_m$ as
\[
\tilde{Q}_m=\left(\begin{array}{ccc}  &  &\left({\bf S_1}^T{\bf
S_m}\right)\\ & &  \left({\bf S_2}^T{\bf S_m}\right)\\
& \tilde{Q}_{m-1}&  ...\\
& &  ...\\
\left({\bf S_m}^T{\bf S_1}\right)&
\left({\bf S_m}^T{\bf S_2}\right)& ...  \left({\bf S_m}^T{\bf
S_m}\right) \end{array}\right)
\]

Now parameterize ${\bf S}_m=r_m\left({\cal O}_N{\bf e}\right)$ as
before, and for $k=1,2,...,m-1$ introduce new vectors $\tilde{\bf
S}_k={\cal O}_N{\bf S}_k$ as integration variables. Obviously, the
entries of the matrix $\tilde{Q}_{m-1}$ do not change, whereas
\[
\left({\bf S_k}^T{\bf S_m}\right)=r_m
\left(\tilde{{\bf S}}^T_k{\cal O}^T_N{\cal O}_N{\bf e}\right)=
r_m\left(\tilde{{\bf S}}^T_k{\bf e}\right)=r_m\tilde{S}_{N,k}
\]
where $\tilde{S}_{N,k}$ stands for the last ($N-$th) component
of the vector $\tilde{{\bf S}}_k$. Further, let us consider
first $N-1$ components of the vector $\tilde{{\bf S}}_k$
as forming the vector ${\bf \zeta}_k$, for the last $N-$th component
of the vector $\tilde{{\bf S}}_k$
using the notation: $\tilde{q}_{k,m}$, $k=1,...,m-1$.
Obviously:
\begin{eqnarray}
d{\bf S}_k&=&d\tilde{{\bf S}}_k=d{\bf \zeta}_k d\tilde{q}_{k,m}\\
\left(\tilde{Q}_{m-1}\right)_{k,l}&=&
\left(\tilde{Q}^{\zeta}_{m-1}\right)_{k,l}+\tilde{q}_{k,m}
\tilde{q}_{l,m}\quad,\quad 1\le k,l\le m-1
\end{eqnarray}
where $\left(\tilde{Q}^{\zeta}_{m-1}\right)_{k,l}=\left(
{\bf \zeta}_k^T{\bf \zeta}_l\right)$

After all those preparations we
change the order of integrations (which is legitimate
in view of the condition Eq.(\ref{conv1}) and the Fubini theorem)
and represent the integral in the right-hand side of Eq.(\ref{defint}) as:
\[
I_{N,M}=\Omega_{N}\int_{R^+}dr r^{N-1}\int_{R^{m-1}}
d\tilde{q}_{1,m}...d\tilde{q}_{m-1,m}
\int_{R^{N-1}}d{{\bf \zeta}_1} ... \int_{R^{N-1}}d{{\bf \zeta}_m}
{\cal F}_1\left[\tilde{Q}_{{\bf \zeta}}\right]
\]
where
\[
F_1\left[\tilde{Q}_{\zeta}\right]=
{\cal F}\left[\left(\begin{array}{ccc}  &  &r_m\tilde{q}_{1,m}
\\ & &  r_m\tilde{q}_{1,m}\\
& \left(\tilde{Q}^{\zeta}\right)_{k,l}+\tilde{q}_{k,m}
\tilde{q}_{l,m}&  ...\\
& &  ...\\
r_m\tilde{q}_{m,1}&
r_m\tilde{q}_{m,2}& ... \, r_m\tilde{q}_{m,m} \end{array}\right)\right]
\]

Now we can apply the equation Eq.(\ref{ind}) to replace the
integration over
the vectors ${\bf \zeta}_k$ to that over the corresponding
positive definite matrices. Further introducing quantities
$q_{k,m}=r_m\tilde{q}_{k_m}\,,\, k=1,...,m-1$ as integration
variables and denoting $r_m^2=q_{m,m}$ we immediately see that
the above integral can be written as:
\begin{eqnarray}\label{A}
&&{\cal C}^{(o)}_{N,1}{\cal C}^{(o)}_{N-1,m-1}\int_{R^+}dq_{m,m}
\int_{R}dq_{1,m}...\int_{R}dq_{m-1,m}
\\ \nonumber &\times& \int_{\hat{Q}_{m-1}>0}d\hat{Q}_{m-1}
q^{(N-n-1)/2}\left[\mbox{det} \hat{Q}_{m-1}\right]^{(N-n-1)/2}
{\cal F}\left[\hat{Q}_{m}\right]
\end{eqnarray}
where we denoted
\[
\hat{Q}_{m}=
\left(\begin{array}{ccc}  &  &q_{1,m}
\\ & &  q_{1,m}\\
& \left(Q_{m-1}\right)_{k,l}+\frac{q_{k,m}q_{l,m}}{q_{m,m}}&  ...\\
& &  ...\\
q_{m,1}& q_{m,2}& ...\,\,  q_{m,m} \end{array}\right)
\]

Using the determinant identity
\[
\mbox{det}\left(\begin{array}{cc}
\hat{Q}_{m-1}& {\bf q}^T\\{\bf q}& q\end{array}\right)
=q\times \mbox{det}\left(\hat{Q}_{m-1}-\frac{{\bf q}\otimes {\bf q}^T}{q}\right)
\]
we see that:
\[
\det{\left[\hat{Q}_{m}\right]}=q_{m,m}\times \det{\left[\hat{Q}_{m-1}\right]}
\]
A little more thinking shows that the conditions
 $q_{m,m}>0$ and $\hat{Q}_{m-1}>0$ ensure that
$\hat{Q}_{m}$ is a general symmetric positive definite of dimension $m$.
For, all minors of the matrix $\hat{Q}_{m}$ either just coincide with
the minors of the matrix $\hat{Q}_{m-1}$ or with those of
the matrix with elements
$Q^{k,l}_{m-1}+\frac{q_{k,m}q_{l,m}}{q_{m,m}}$, both being positive
definite.

Combining all this knowledge and the fact that
${\cal C}^{(o)}_{N,1}{\cal C}^{(o)}_{N-1,m-1}={\cal C}^{(o)}_{N,m}$
we see that the formula
Eq.(\ref{A}) can be written exactly as the right-hand side of the
equation Eq.(\ref{trans1}) thus completing the proof.

 Theorem I then follows by trivially repeating, mutatis mutandis,
all the steps of the proof given above for the case of complex
vectors and Hermitian matrices, replacing the orthogonal matrices
with unitary one as appropriate.




\begin{thebibliography}{99}

\bibitem{zeta1} JP Keating , NC Snaith "Random Matrix Theory and
$\zeta(1/2+it)$", Comm. Math. Phys., {\bf 214} (2000), 57;
\bibitem{zeta2} JP Keating, NC Snaith
"Random matrix theory and L-functions at s=1/2" Comm. Math. Phys.
{\bf 214},(2000) 91
\bibitem{zeta3} CP Hughes, JP Keating, N O'Connell
"Random matrix theory and the derivative of the Riemann zeta
function" P Roy Soc Lond A Mat {\bf 456} (2000) 2611
\bibitem {BH} E. Brezin and S. Hikami , "Characteristic Polynomials
of Random Matrices", Comm. Math. Phys., {\bf 214}, (2000), 111-135
and "Characteristic Polynomials of Random Real Symmetric Matrices"
Comm. Math. Phys., {\bf 223}, (2001), 363-382
\bibitem{web} Workshop "L-functions and Random Matrix Theory", The American
Institute of Mathematics, www.aimath.org/PWN/Irmt/index.html
\bibitem{Gan} DM Gangardt , "Second Quantization approach to
characteristic polynomials in RMT", J. Phys.A: Math.Gen. {\bf 34}
(2001) 3553
\bibitem{GK} DM Gangardt and A Kamenev,
"Replica treatment of the Calogero-Sutherland model",
 Nucl. Phys. B, {\bf 610} (2001),578 (e-preprint arXiv:cond-mat/0102405)
\bibitem{U1} F.Haake, M.Kus, H.-J.Sommers, H.Schomerus, K.Zyckowski,
 "Secular determinants of random unitary matrices",
J.Phys.A: Math.Gen., {\bf 29} (1996), 3641
\bibitem{U2} S Ketteman, D.Klakow and U. Smilansky "Characterization of
quantum chaos by the autocorrelation function of spectral
determinants" J.Phys.A: Math.Gen., {\bf 30} (1997), 3643
\bibitem{F} Y V Fyodorov "Spectra of Random Matrices Close to Unitary
and Scattering Theory for Discrete-Time systems", in: "Disordered
and Complex Systems", edited by P.Sollich et al., AIP Conference
Proceedings v.553, Melville NY, 2001
\bibitem{MN} ML Mehta, J-M Normand, "Moments
of the characteristic polynomial in the three ensembles of random
matrices" ,  J.Phys A.Math.Gen: {\bf 34} (2001), 4627
\bibitem{KM} A Kamenev and M Mezard, "Wigner-Dyson statistics from
the Replica Method", J Phys.A, {\bf 32} (1999) 4373
\bibitem{LY} I.V.Yurkevich and I.V.Lerner, "Nonperturbative results for level correlations
from the replica nonlinear sigma model," Phys.Rev.B {\bf 60}, 3955
\bibitem{AS} A V Andreev and BD Simons, "Correlator
of the Spectral Determinants in Quantum Chaos",  Phys.Rev.Lett.
{\bf 75} (1995), 2304
\bibitem{FK99}
Y V Fyodorov and B A Khoruzhenko, "Systematic Analytical Approach
to Correlation Functions of Resonances in Quantum Chaotic
Scattering". Phys. Rev. Lett. {\bf 83} (1999), 66
\bibitem{AGG} A Cavagna, J Garrahan, I Giardina, " Index Distribution
of random matrices with an Application to Disordered Systems",
 Phys. Rev. B {\bf 61} (2000),  3960
\bibitem{TS} T Shirai, " A Factorization of Determinant Related to
Some Random Matrices", J. Stat. Phys. {\bf 90} (1998), 1449
\bibitem{Ket}
S Kettemann  "Exploring level statistics from quantum chaos to
localization with the autocorrelation function of spectral
determinants" Phys. Rev. B {\bf 59} (1999), 4799 and S Kettemann
and A Tsvelik "Information about the integer quantum Hall
transition extracted from the autocorrelation function of spectral
determinants" , Phys. Rev. Lett. {\bf 82 } (1999) 3689
\bibitem{I} YV Fyodorov "Negative moments of characteristic
polynomials of random matrices: Ingham-Siegel integral as an
alternative to Hubbard-Stratonovich transformation",
 Nucl. Phys. B {\bf B}[PM] (2001) 621 (e-preprint arXiv:math-ph/0106006).
\bibitem{ZN}
S Nonnenmacher an MR Zirnbauer, "Det-Det Correlations for quantum
maps, dual pair and saddle-point analysis", e-preprint
arXiv:math-ph/0109025.
\bibitem{DN} PH Damgaard, SM Nishigaki "Universal spectral
correlators and massive Dirac operators",
 Phys.Rev.B, {\bf 57} (1998), 5299
\bibitem{BK} MV Berry and JP Keating, "Clasters of near-degenerate
levels dominate negative moments of sectral determinants", J.
Phys.A: Math.Gen. {\bf 35} (2002) L1-L6
\bibitem{E} E Strahov "Moments of characteristic polynomials
enumerate lexicographic arrays", e-preprint arXiv:math-ph/0112043
\bibitem{IZ} C.Itzykson and J.B.Zuber,
"The planar approximation. II", J.Math.Phys. {\bf 21} (1980), 411
\bibitem{HC} Harish-Chandra , "Differential operators on a semisimple
Lie algebra", Proc.Nat.Acad.Sci. {\bf 42}, (1956) 252
\bibitem{DH} JJ Duistermaat, GJ Heckman "On the variation
in the co-homology of the symplectic form of the reduced phase
space " "Invent.Math. {\bf 69} (1982), 259  and {\it ibid} 1983
 Invent. Math. {\bf 72} 153
\bibitem{SW} L Sch\"{a}fer and F Wegner, "Disordered System with $n$
Orbitals per Site: Lagrange formulation, Hyperbolic Symmetry, and
Goldstone modes", Z. Physik B-Condensed Matter, {\bf 38} (1980),
113
\bibitem{Farmer} D W Farmer "Long mollifiers of the Riemann
zeta-function",  Mathematika, {\bf 40} (1993), 71
\bibitem{Efetov} K.B. Efetov, "Supersymmtry in Disorder and Chaos"
(Cambridge University Press, Cambridge 1997).
\bibitem{VZ} JJM Verbaarschot and MR Zirnbauer, "Critique of the
Replica Trick", J Phys. A:Math.Phys., {\bf 17} (1985), 1093
\bibitem{anom} MJ Rothstein "Integration on noncompact
supermanifolds" Trans. Amer. Math. Soc.
{\bf 299} (1987), 387
\bibitem{Zirna} M R Zirnbauer, "Supersymmetry for systems
with unitary disorder: circular ensembles" J Phys. A:Math.Phys., {\bf 29}
(1996), 7113
\bibitem{Szabo1} RJ Szabo "Microscopic Spectrum of the QCD Dirac
Operator in Three Dimensions", Nucl. Phys.B, {\bf 598} (2001), 309
\bibitem{Guhr} T Guhr " Dyson Correlation functions and Graded Symmetry",
J.Math.Phys. {\bf 32}, 336 (1991)
\bibitem{prato}
E Prato and  S Wu, "Duistermaat-Heckman measures in a non-compact
setting", arXiv:alg-geom/9307005 , Compos. Math. {\bf{94}} (1994)
113
\bibitem{rossmann}
W. Rossmann, "Kirilov's character formula for reductive group",
Invent. Math. (1978) {\bf 48} 207
\bibitem{berline}
N. Berline and M.Vergne, "Fourier transform of orbits of the
coadjoint representation" in {\it Representation theory of
reductive groups},  (1983) pp.53-57, Birkh\"auser, Basel
\bibitem{paradan}
P.-E. Paradan, "The Fourier Transform of Semi-Simple Coadjoint
Orbits" { \it J Funct Anal} {\bf 163} (1999) 152
\bibitem{Zirn} M R Zirnbauer, "Another Critique of the Replica
Trick", e-preprint arXiv:cond-mat/9903338
\bibitem{bordemann}
M Bordemann, M Forger, and H R\"{o}mer " Homogeneous K\"ahler
Manifolds: Paving the Way Towards New Supersymmetric Sigma Models"
 Commun. Math. Phys. (1986) {\bf 102} 605
\bibitem{helgason}
S Helgason  2000 {\it Differential Geometry and Symmetric Spaces}
(AMS Chelsea Publishing)
\bibitem{IS} A E Ingham, "An integral which Occurs in
Statistics",  Proc.Camb.Phil.Soc.,{\bf 29}(1933), 271;
 C L Siegel, "\"{U}ber der analytische Theorie der
quadratischen Formen", Ann. Math. {\bf 36}(1935), 527
\bibitem{Hua}
L.K. Hua, {\it Harmonic Analysys of Functions of Several Complex
Variables in the Classical Domains} (AMS, Providence, 1963).
\bibitem{szabo}
R J Szabo "Equivariant localization of path integrals",
e-preprint arXiv:hep-th/9608068
\bibitem{picken}
R F Picken,  "The Duistermaat-Heckman integration formula on flag
manifolds",  J. Math. Phys. (1990) {\bf 31}(3) 616
\bibitem{borel}
A Borel " K\"{a}hlerian coset spaces of semisimple Lie groups"
 Proc. Natl. Acad. Sci. USA  (1954){\bf 40}, 1147
\bibitem{marinov1}
D Bar-Moshe and M S Marinov "Realization of compact Lie
algebras in K\"{a}hler manifolds" {\it J. Phys. A} {\bf 27} (1994), 6287
\bibitem{marinov2}
D Bar-Moshe and  M S Marinov, "Berezin quantization and unitary
representations of Lie groups," in {\it Berezin Memorial}, edited
by R. Dobrushin, M. Shubin, and A. Vershik (American Mathematical
Society, Providence, RI, 1995)
\bibitem{kobayashi}
S Kobayashi and K Nomizu {\it Foundations of Differential
Geometry}  (Interscience, New-York, 1969) Vol. 2, Chap. 9
\bibitem{arnold}
V I Arnold {\it Mathematical Methods of Classical
Mechanics} (Springer, 1978)
\bibitem{bismut}
J M Bismut "Localisation formulas, Superconnections, and the Index theorem for
families"  Commun. Math. Phys. {\bf 103} (1986) 127
\bibitem{witten}
E Witten
"Two Dimensional Gauge Theories Revisited" (ArXiv:hep-th/9204083)
 {\it J. Geom. Phys.} {\bf 9} (1992), 303
\bibitem{bando1}
M Bando , T Kuramoto, T Maskawa and S Uehara
 "Structure of non-linear realization in supersymmetric theories"
{\it Phys. Lett.} {\bf{B}} 138 (1984), 94
\bibitem{itoh}
K Itoh , T Kugo and H Kunitomo,  "Supersymmetric nonlinear
realization for arbitrary K\"{a}hlerian coset space ${\sf{G/H}}$"
{\it Nucl. Phys.} {\bf B} 263(1986) , 295
\bibitem{fujii}
K Fujii and K Funahashi "Multi-Periodic Coherent States and
WKB-Exactness II" {\it J. Math. Phys} {\bf 38}(1997), 2812
\bibitem{Dupl} F David, B Duplantier and E Guitter,
 " Renormalization Theory for Interacting Crumpled Manifold",
Nucl.Phys.B, {\bf 394} (1993), 555
\bibitem{guhr}
T. Guhr and T. Wettig "An
Itzykson-Zuber-like integral and diffusion for complex ordinary
and supermatrices", {\it J Math Phys} {\bf 37}(12) (1996) 6395
\bibitem{verbaarschot}
A. D. Jackson, M. K. Sener and J. J. M. Verbaarschot, "Finite
volume partition functions and Itzykson-Zuber integrals", {\it
Phys. Lett.} {\bf B} 387 (1996) 355
\end{thebibliography}
\end{document}